\documentclass[
  journal=pasa,
  manuscript=research-paper, 
  year=202X,
  volume=YY,
]{cup-journal}

\usepackage{amsmath}
\usepackage[nopatch]{microtype}
\usepackage{hyperref}
\usepackage{aas-macros}
\usepackage{multirow}
\usepackage{amssymb,siunitx,booktabs}
\hypersetup{
    colorlinks=true,
    linkcolor=blue,
    filecolor=magenta,      
    urlcolor=cyan,
    pdftitle={Overleaf Example},
    pdfpagemode=FullScreen,
    }

\title[]{Enhanced Astrometry of the Rapid ASKAP Continuum Survey: Mid and High Frequency Epochs}

\author{Akhil Jaini}
\affiliation{Center for Astrophysics and Supercomputing, Swinburne University of Technology, Post Office Box 218, Hawthorn, VIC 3122, Australia}
\email[Akhil Jaini]{ajaini@swin.edu.au}

\author{Adam T. Deller}
\affiliation{Center for Astrophysics and Supercomputing, Swinburne University of Technology, Post Office Box 218, Hawthorn, VIC 3122, Australia}
\alsoaffiliation{ARC Centre of Excellence for Gravitational Wave Discovery (OzGrav), Post Office Box 218, Hawthorn, VIC 3122, Australia}

\author{Yuanming Wang}
\affiliation{Center for Astrophysics and Supercomputing, Swinburne University of Technology, Post Office Box 218, Hawthorn, VIC 3122, Australia}
\alsoaffiliation{ARC Centre of Excellence for Gravitational Wave Discovery (OzGrav), Post Office Box 218, Hawthorn, VIC 3122, Australia}

\author{Emil Lenc}
\affiliation{CSIRO Space and Astronomy, Post Office Box 76, Epping, NSW 1710, Australia}

\author{Stefan W. Duchesne}
\affiliation{CSIRO Space and Astronomy, Post Office Box 1130, Bentley, WA 6102, Australia}

\author{Marcin Glowacki}
\affiliation{Institute for Astronomy, University of Edinburgh, Royal Observatory, Edinburgh, EH9 3HJ, United Kingdom}
\alsoaffiliation{Inter-University Institute for Data Intensive Astronomy, Department of Astronomy, University of Cape Town, Cape Town, South Africa}

\received {dd Mmm YYYY}
\revised  {dd Mmm YYYY}
\accepted {dd Mmm YYYY}
\published{22 September 202X}


\begin{document}

\begin{abstract}
Accurate radio astrometry is essential for reliable cross-identification of sources across wavelengths, precision localisation of transient events, and the construction of stable all-sky reference catalogues. In this work we extend our astrometric correction framework for the Rapid ASKAP Continuum Survey (RACS) to its mid- and high-frequency epochs (RACS-Mid1 and RACS-High1), building on our previous corrections to the low-frequency surveys. Using a hierarchical crossmatching strategy with high-precision external catalogues, we remove large-scale systematic positional errors that were present in the uncorrected data and significantly reduce the residual scatter across the sky. After correction, the median positional offsets are effectively eliminated, and the 68\% confidence interval of the mean residuals of RACS source positions averaged over $~\sim1$~sq. deg. regions is reduced from $\gtrsim 0.4''$ to $\lesssim 0.18''$ over most of the survey area for both epochs. Independent validation against multiple external radio astrometric references confirms that individual corrected RACS source positions are accurate to a 1-$\sigma$ confidence level of $\sim 0.25''$ across the bulk of the sky, with slightly degraded performance within the Galactic plane. While motivated primarily by the need for improved localisation of ASKAP fast radio bursts, these corrections also benefit a wide range of science applications, including transient identification, multiwavelength host association, and studies of Galactic and extragalactic radio populations. Together with our previous work, this establishes RACS as the highest precision arcsecond-resolution all-sky astrometric reference in the southern hemisphere at decimetre wavelengths.
\end{abstract}

\section{INTRODUCTION} \label{sec:introduction}

The Australian Square Kilometre Array Pathfinder (ASKAP) is an interferometer located in Western Australia. It is made up of 36 antennas of 12~m diameter, working in the frequency range of 700-1800~MHz, with a maximum baseline length of 6~km. Each antenna is equipped with novel Phased-Array Feed (PAF) receivers that, together with advanced digital systems, generate 36 simultaneous, independent beams, which collectively cover a $\sim30$~sq.deg. field-of-view (\citealp{2021PASA...38....9H}). The Rapid ASKAP Continuum Survey (RACS, \citealp{2020PASA...37...48M}) provides the first arcsecond-scale view of the Southern radio sky, delivering extensive coverage of the sky across ASKAP’s full frequency range. ASKAP itself is unique in combining a very wide field-of-view with arcsecond-scale resolution, making it one of the few instruments in the southern hemisphere capable of producing a survey of this breadth and fidelity. RACS therefore provides a foundation for a wide range of science, from the study of radio-loud active galactic nuclei and star-forming galaxies (\citealp{2021A&A...647L..11I}, \citealp{2021Galax...9...99A}, \citealp{2022MNRAS.511.3525D}), to probing the time-variable radio sky (\citealp{2022NatAs...6..828C}, \citealp{2023ApJ...948...67B}, \citealp{2023ApJ...956...28A}). The value of this resource, however, depends significantly on the accuracy of its astrometry, as precise source positions are required for robust cross-identification with surveys at other wavelengths, construction of broadband spectral energy distributions, and reliable population studies (\citealp{2021PASA...38...50D}, \citealp{2023MNRAS.523.5661W}).


In our preceding work \cite{jaini2025enhanced}, we developed and implemented a methodology for correcting systematic astrometric errors in the low-frequency epochs of RACS (RACS-Low1 and RACS-Low3, \citealp{2021PASA...38...58H}). These corrections reduced large-scale positional biases to effectively zero and achieved typical $1-\sigma$ uncertainties of $\sim0.3''$ in most regions of the sky, increasing only slightly within the Galactic plane. This represented a substantial advance over the native catalogues, enabling sub-arcsecond localisation of fast radio bursts (FRBs) detected by ASKAP, while also enhancing the utility of RACS for broader astrophysical studies.

While FRBs and other radio transients remain primary drivers of this work, the benefits of precise astrometry extend much further. Improved RACS positions enable robust associations with crowded optical and infrared catalogues, critical for the study of flare stars, ultra-long-period radio transients, and other rare phenomena (\citealp{2025A&A...698A.158I}, \citealp{2024PASA...41....4B}). High-precision astrometry also underpins the accurate classification of radio galaxies, improves the fidelity of spectral and morphological studies, and provides a reference frame for future southern hemisphere surveys. In the northern hemisphere, multiwavelength and transient studies have long benefited from the robust astrometric framework provided by large-area radio reference catalogues such as the NRAO VLA Sky Survey (NVSS, \citealp{1998AJ....115.1693C}) and, more recently, the Very Large Array Sky Survey (VLASS, \citealp{2020PASP..132c5001L}). But in the southern hemisphere, the only comparable survey was the Sydney University Molonglo Sky Survey (SUMSS, \citealp{1999AJ....117.1578B}), which has much worse astrometry and resolution. With the corrections presented here, RACS now extends wide-field, sub-arcsecond radio astrometric capability to the southern sky, comparable to its northern hemisphere counterparts, ensuring that precision radio reference catalogues are available across the whole sky.



Building on our previous research, this paper focuses on establishing a uniform, high-precision astrometric reference across the full ASKAP frequency range by enhancing the astrometry of the RACS-Mid1 epoch (central frequency 1367.5 MHz, \citealp{2023PASA...40...34D}) and the RACS-High1 epoch (central frequency $\sim 1655.5$ MHz, \citealp{duchesne2025rapid}). Here we extend that framework to the higher-frequency surveys while incorporating lessons learned from the RACS-Low3 analysis (as detailed in Section 2.2 of \citealp{jaini2025enhanced}), particularly regarding how catalogue construction and sky-dependent systematics influence astrometric stability. We introduced a refined correction strategy that is more robust in regions of high source density and better adapted to variable observing conditions. The updated approach uses an improved crossmatching hierarchy and a more flexible offset modelling scheme designed to reduce false associations and preserve statistical precision in challenging parts of the sky. Together, these changes enable more consistent astrometric performance across the entire RACS sky coverage.


The paper is structured as follows: Section~\ref{sec:methodology} describes the refined methodology employed for astrometric correction; Section~\ref{sec:results} presents the quantitative results for RACS-Mid1 and RACS-High1, including internal validation and external cross-checks; Section~\ref{sec:discussion} explores the implications for a range of scientific applications; and Section~\ref{sec:conclusion} summarises our findings and outlines future directions.

\begin{figure*}[h!]
    \centering
    \includegraphics[scale=0.53]{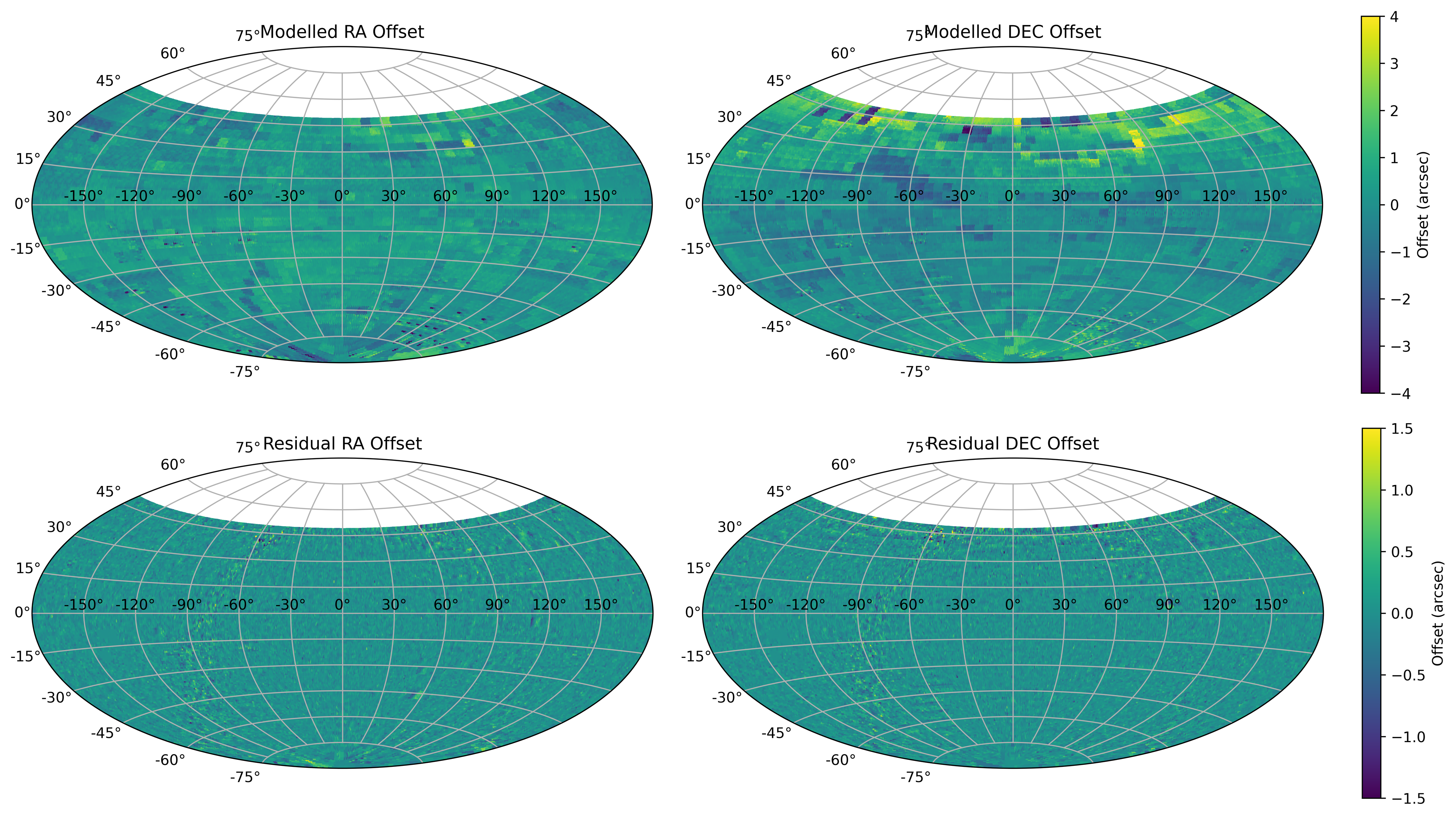}
    \caption{The modelled and residual offsets of RACS-Mid1 vs WISE after two stages of crossmatching are shown here. The top row shows the modelled offsets and the bottom row shows the residual offsets in RA (left) and Dec. (right) for the entire sky coverage.}
    \label{fig:racsmid1aitoff}
\end{figure*}

\begin{figure*}[h!]
    \centering
    \includegraphics[scale=0.65]{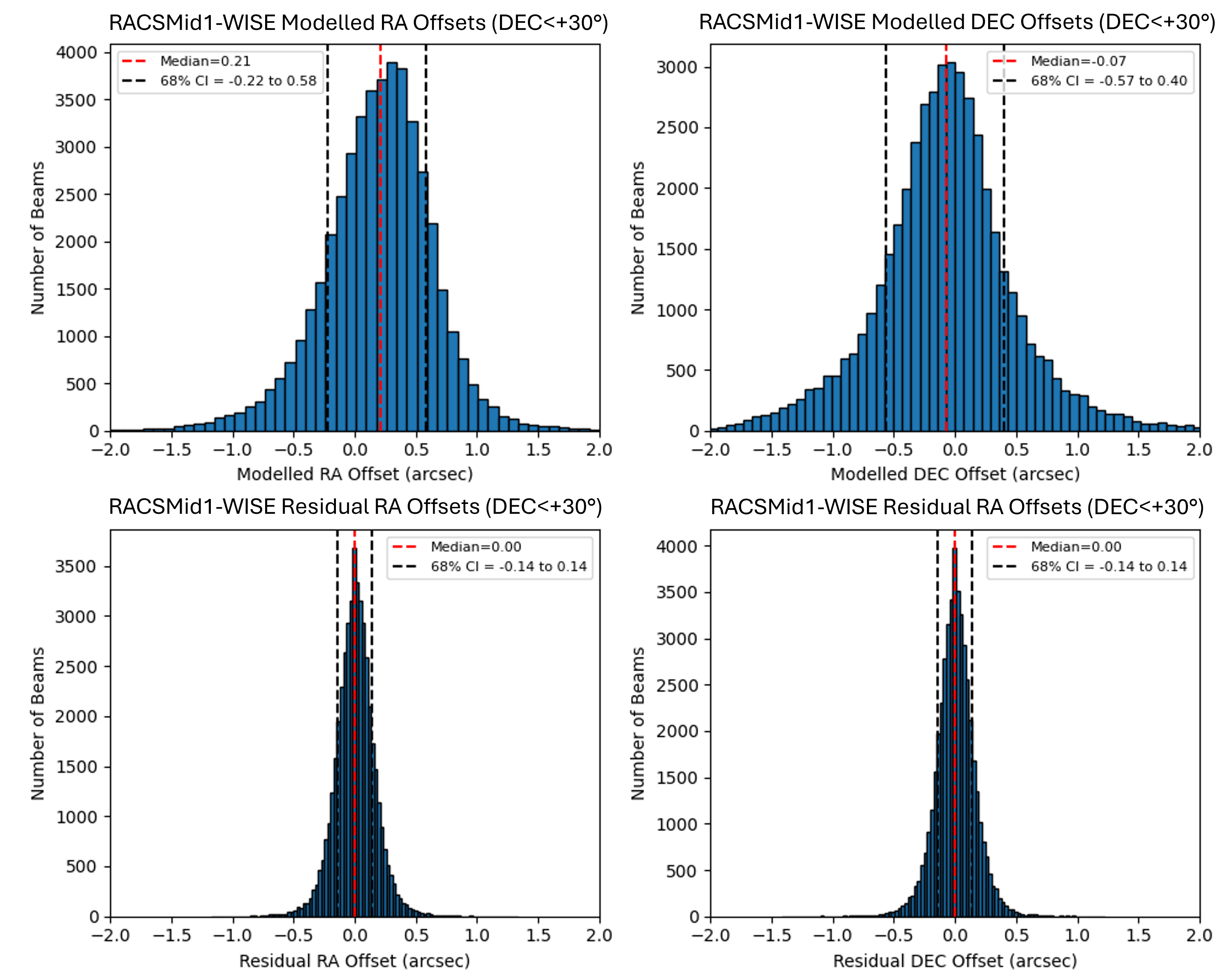}
    \caption{The modelled (top row) and residual (bottom row) offsets for RACS-Mid1 vs WISE for scans below Dec. $+30^\circ$. The median RA offset (left) improves from $0.21''$ to $0.00''$ and the median Dec. offset (right) decreases from $-0.07''$ to $0.00''$. The 68\% confidence intervals also narrow down significantly.}
    \label{fig:racsmid1histograms}
\end{figure*}

\section{METHODOLOGY} \label{sec:methodology}

This section covers changes to the original approach described in \citealp{jaini2025enhanced}. The entire modelling pipeline\footnote{GitHub repository: \url{github.com/jainiakhil/RACS_Astrometry}} was developed in Python version 3.11, along with relevant modules such as \textit{astroquery} (\citealp{2019AJ....157...98G}), \textit{astropy} (\citealp{astropy:2013}, \citealp{astropy:2018}, \citealp{astropy:2022}), \textit{scipy} (\citealp{2020SciPy-NMeth}), and \textit{numpy} (\citealp{harris2020array}).

The primary aim of this work is to derive accurate astrometric corrections for every source in the RACS-Mid1 and RACS-High1 epochs. To maintain consistency across ASKAP’s frequency bands, we adopted a framework that closely follows the procedure developed for RACS-Low1 and RACS-Low3 \citep{jaini2025enhanced}, while introducing several important refinements motivated by the limitations we encountered in those earlier studies. Below, we first summarise the common workflow before outlining the specific changes implemented for these higher-frequency epochs.

The RACS survey strategy is based on ASKAP’s phased-array feed receivers, which form 36 simultaneous primary beams on the sky. Each observation of a given field is referred to as a \textit{scan}, and every scan consists of a $6 \times 6$ grid (\textit{closepack36} footprint, \citealp{2021PASA...38....9H}) of beams covering approximately 30 square degrees of sky. The individual ASKAP beams from each scan are normally combined into larger mosaicked images, from which the public RACS catalogues are generated to cover the full survey area. In this work, however, we operate directly on the per-beam catalogues derived from the individual beam images, which we refer to as \textit{source lists}. This choice avoids positional systematics that can arise near beam boundaries during mosaicking, where blended or partially overlapping beam responses can bias centroid measurements. Working at the per-beam level therefore preserves the native astrometric information of each observation and provides a cleaner basis for modelling systematic offsets. Thus, each \textit{beam} represents the fundamental unit of analysis, and a \textit{scan} comprises the collection of beams observed simultaneously during a single pointing. It can also be noted that these per-beam source lists are not primary beam corrected, which is usually a step undertaken during the mosaicking process, and therefore would have apparent brightness/flux densities that are attenuated by the primary beam.


For every beam, we defined a circular region of radius $1^\circ$ centred on the nominal beam position to provide sufficient crossmatches. Each scan was then processed independently, allowing us to model systematic calibration errors across both beam and scan levels. As in our earlier work, the Widefield Infrared Survey Explorer (WISE, \citealp{2010AJ....140.1868W}) all-sky infrared catalogue was adopted as the primary astrometric reference due to its complete coverage, sub-arcsecond accuracy, and high source density. 


\begin{figure}[h!]
    \centering
    \includegraphics[scale=0.65]{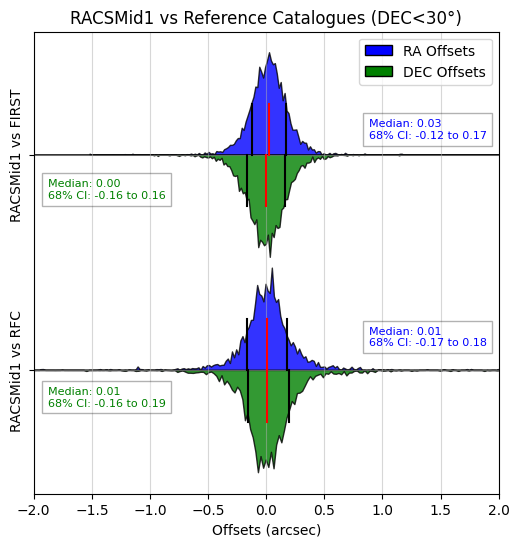}
    \caption{The corrected offsets in RA and Dec. for Dec. < $+30^\circ$ for RACS-Mid1 vs reference catalogues, with FIRST comparisons in the top row and RFC in the bottom row.}
    \label{fig:racsmid1firstrfc}
\end{figure}

\subsection{Common Framework}
The overall correction workflow can be summarised as follows:
\begin{enumerate}

    \item Partition the data into individual ASKAP beams (each $1^\circ$ in radius centred on the nominal beam position).
    \item Apply an initial source filtering step: retain only RACS and WISE sources with SNR $>6$, exclude extended or blended RACS sources (the ratio of integrated to peak flux $>1.5$, or multiple Gaussian components within $30''$).
    \item Perform two-stage crossmatching:
    \begin{itemize}
        \item Stage~1: crossmatch against a reference catalogue to capture large-scale offsets (crossmatching thresholds are defined in Section~\ref{sec:newrefinements});
        \item Stage~2: refine corrections with WISE using a $2''$ threshold.
    \end{itemize}
    \item Model calibration-induced offsets as the sum of two terms:
    \begin{itemize}
        \item \textit{Scan-independent beam offset}: common to all scans sharing a bandpass calibration;
        \item \textit{Beam-independent scan offset}: common to all beams within a given scan.
    \end{itemize}
    \item Apply the derived corrections to positions of the catalogue sources or the source lists, and estimate beam-wise residual uncertainties with 68\% and 95\% confidence intervals.
    \item Validate the corrections against independent reference catalogues (further described in Section~\ref{sec:newrefinements}).
\end{enumerate}

\begin{figure*}[h!]
    \centering
    \includegraphics[scale=0.53]{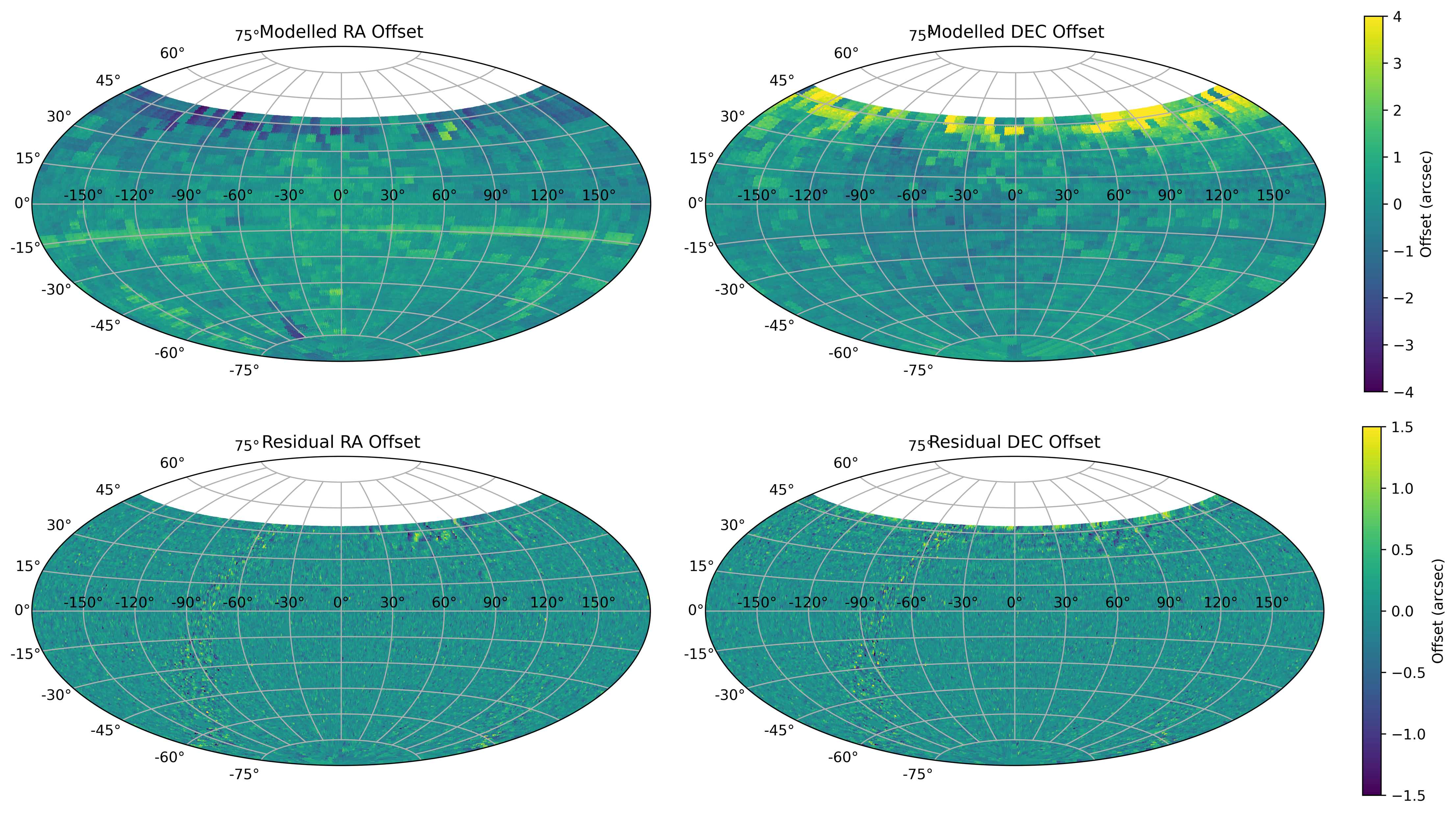}
    \caption{The modelled and residual offsets of RACS-High1 vs WISE after two stages of crossmatching are shown here. The top row shows the modelled offsets and the bottom row shows the residual offsets in RA (left) and Dec. (right) for the entire sky coverage.}
    \label{fig:racshigh1aitoff}
\end{figure*}

\subsection{Refinements for RACS-Mid1 and RACS-High1} \label{sec:newrefinements}
While the overall workflow remains unchanged, several key refinements were introduced to improve reliability at mid and high frequencies. From our previous analysis of RACS-Low3, we recognised that offsets in the Galactic plane and at higher declinations often exceeded the standard crossmatching radius of $5''$, with some beams showing offsets as large as $12''$. Attempting to correct these regions directly with WISE at such large radii would have resulted in an unacceptable rate of false-positive matches, owing to the very high density of WISE sources of approximately 13,500 sources per square degree on average. 


To address this, we modified the first stage of the crossmatching process. Instead of beginning with WISE, we crossmatched the RACS-Mid1 and RACS-High1 beams against the corrected RACS-Low1 positions (for Dec.\ $<0^\circ$) and against VLASS (for Dec.\ $>0^\circ$). Both of these surveys have an average source density of approximately 100 sources per square degree, substantially lower than that of WISE, and both provide astrometric accuracy at the $\sim0.1''$--$0.3''$ level. This ensured that the initial alignment between catalogues was itself highly reliable, and that the subsequent $2''$ crossmatching radius used for WISE remained conservative and safe. By leveraging catalogues with comparable angular resolution, source density and frequency coverage to the mid- and high-band ASKAP data, we were able to extend the initial crossmatching radius to $12''$ to capture the larger offsets accurately but still with a lower false positive rate. RACS-Low1 was selected over RACS-Low3 because of the greater stability and reliability of its corrections, while VLASS was adopted for northern declinations where RACS-Low1 becomes less reliable. For this crossmatching, we used the VLASS quick look epoch 1 catalogue of the entire sky (\citealp{2021ApJS..255...30G}). This hierarchical approach provided a robust mechanism for recovering large-scale offsets while maintaining strict control over spurious associations.

The second stage of corrections remained unchanged, relying on a $2''$ crossmatching radius with WISE to refine the astrometry. After filtering and applying both per-scan and per-beam corrections, we measured the residuals against WISE to estimate final uncertainties. These results were further validated against VLA-FIRST (the Very Large Array - Faint Images of the Radio Sky at Twenty-cm survey, \citealp{1994ASPC...61..165B}) and RFC (the Radio Fundamental Catalog, \citealp{petrov2025radio}), though not against VLASS since it was now used during the first stage of crossmatching. The comparison with RFC measures the residual offset on a per-source basis, however, since RFC positions are determined using Very Long Baseline Interferometry (VLBI) with milliarcsecond resolution, any intrinsic arcsecond-scale structure resolved by ASKAP but not by VLBI will manifest as an additional apparent offset. Consequently, the RFC-based residuals should be interpreted as a conservative upper limit on the true RACS astrometric uncertainty.

At higher frequencies, particularly in the RACS-High1 epoch, we also considered the impact of intrinsic centroid differences between the infrared and radio bands. The higher spatial resolution of RACS-High1 meant that in some cases it resolved structure not seen in WISE, which led to broader residual distributions. While this effect did not introduce systematic biases into the mean corrections, it did increase the observed scatter in the residuals. To account for this, we based our final astrometric uncertainty estimates on the empirically measured 68\% confidence intervals of the RACS--WISE residual offset distributions (as presented in Section~\ref{sec:results}), rather than relying solely on the formal positional uncertainties reported in the catalogues. This ensures that the quoted astrometric precision captures both measurement noise and the additional scatter introduced by intrinsic centroid differences between radio and infrared observations.


Finally, unlike in RACS-Low3, we did not exclude the Galactic plane or regions above Dec.\ $+30^\circ$ when constructing the offset model. In earlier work this exclusion was introduced in an attempt to improve modelling, but the gains proved marginal. With the new Stage~1 crossmatching against RACS-Low1 and VLASS, the additional filtering became unnecessary. These regions are still more challenging due to source density and degraded observing conditions, and users should exercise caution when interpreting results there, but they remain part of the global correction model presented in this work.

\section{RESULTS} \label{sec:results}

Building upon the corrections developed for RACS-Low1 and RACS-Low3, we extended our refined workflow to the mid- and high-frequency epochs of RACS. Both RACS-Mid1 and RACS-High1 consist of 1493 unmosaicked scans comprising 53,748 beams in the same beam footprint as RACS-Low3 (as detailed in \citealp{jaini2025enhanced}). Below we present the results for each epoch in turn.

\subsection{RACS-Mid1 Corrections} \label{sec:racsmid1corrections}

The modelled and residual positional offsets in RACS-Mid1 relative to WISE are displayed in Aitoff projection in Figure~\ref{fig:racsmid1aitoff}. The modified methodology delivers improved corrections across most of the sky, though we continue to recommend caution for Dec. above $+30^\circ$, where the very low observing elevation generally means that simple calibration extrapolation leads to large residual errors, causing the ASKAP performance to degrade. Accordingly, we first quantify the results for the region below Dec.\ $+30^\circ$. 

\begin{table*}[h!]
\centering
\caption{The residual offsets for RACS-Mid1 vs WISE for scans below Dec. $+30^\circ$.}
\begin{tabular}{|c|cc|cc|cc|}
\hline
\textbf{RACS-Mid1--WISE Offsets}   & \multicolumn{2}{c|}{\textbf{All-Sky}}                & \multicolumn{2}{c|}{\textbf{On-Plane}}               & \multicolumn{2}{c|}{\textbf{Off-Plane}}              \\ \cline{2-7} 
\textbf{(Dec < $+30^\circ$, arcsec)}   & \multicolumn{1}{c|}{\textbf{RA}}    & \textbf{DEC}   & \multicolumn{1}{c|}{\textbf{RA}}    & \textbf{DEC}   & \multicolumn{1}{c|}{\textbf{RA}}    & \textbf{DEC}   \\ \hline
Median                      & \multicolumn{1}{c|}{0.00}           & 0.00           & \multicolumn{1}{c|}{0.00}           & 0.00           & \multicolumn{1}{c|}{0.00}           & 0.00           \\ \hline
68\% Confidence Intervals   & \multicolumn{1}{c|}{-0.14 --- 0.14} & -0.14 --- 0.14 & \multicolumn{1}{c|}{-0.20 --- 0.22} & -0.20 --- 0.20 & \multicolumn{1}{c|}{-0.14 --- 0.14} & -0.13 --- 0.13 \\ \hline
95.4\% Confidence Intervals & \multicolumn{1}{c|}{-0.34 --- 0.34} & -0.34 --- 0.33 & \multicolumn{1}{c|}{-0.50 --- 0.48} & -0.48 --- 0.47 & \multicolumn{1}{c|}{-0.31 --- 0.31} & -0.32 --- 0.31 \\ \hline
99.7\% Confidence Intervals & \multicolumn{1}{c|}{-0.66 --- 0.68} & -0.66 --- 0.65 & \multicolumn{1}{c|}{-0.83 --- 0.91} & -0.90 --- 0.86 & \multicolumn{1}{c|}{-0.57 --- 0.60} & -0.63 --- 0.61 \\ \hline
\end{tabular}
\label{tab:racsmid_specs}
\end{table*}

As shown in the histograms in Figure~\ref{fig:racsmid1histograms}, the mean bias is eliminated in both RA and Dec., with the stochastic scatter reduced significantly. The remaining results are summarised in Table~\ref{tab:racsmid_specs}. It is important to note that the residual error distributions are not strictly Gaussian and exhibit longer tails than expected for a normal distribution, as demonstrated quantitatively by the results presented here. The results also show that both the scatter and the degree of non-Gaussianity increase within the Galactic plane, indicating that our corrections are less effective in this region.


When extended to the full RACS-Mid1 sky, including regions above Dec.\ $+30^\circ$, we again find that the mean bias is removed and scatter reduced: the 68\% confidence intervals shrink from $0.20''^{+0.60}_{-0.26}$ to $0.00''^{+0.15}_{-0.15}$ in RA and from $-0.02''^{+0.65}_{-0.55}$ to $0.00''^{+0.15}_{-0.15}$ in Dec. While modest degradations persist in areas with low crossmatch density or high source confusion, the algorithm generalises well to the mid-frequency data overall.  

We then validated the corrected RACS-Mid1 catalogue against FIRST and RFC, combining the comparisons into a single analysis (as shown in Figure~\ref{fig:racsmid1firstrfc}). The verification procedure followed the approach detailed in \citet{jaini2025enhanced}, with some modifications described below, and with the exception that VLASS was not used here, since it had already served as part of the Stage~1 corrections and would therefore bias the test.  

Crossmatching RACS-Mid1 with FIRST, which is only available outside the Galactic Plane, in the same beam-wise bins used for WISE resulted in median post-correction offsets of $0.02''$ in RA and $0.00''$ in Dec., based on $\sim$9,000 overlapping beams with an average of 50 crossmatched sources per beam (within a $1^\circ$ radius). This sample size is sufficient to average down statistical noise and mitigate source-to-source centroid differences arising from intrinsic structure, while still probing spatial scales comparable to a single beam coverage. As a result, the analysis should largely preserve genuine sky-dependent variations in the residual offsets rather than smoothing over real astrometric residuals. 


When the RACS-Mid1 sources, filtered to remove extended and faint sources using the same criteria defined in Section~\ref{sec:methodology}, are crossmatched individually with the high-precision RFC catalogue using a matching radius of $5''$, we obtain median offsets of $0.01''$ in both RA and Dec.\ for approximately 9,000 crossmatches across the entire RACS sky. However, because the higher-frequency RACS-High1 epoch naturally contains a larger fraction of resolved sources than RACS-Mid1 due to its higher angular resolution, some sources that pass the filtering criteria in RACS-Mid1 may be rejected in RACS-High1 once their extended structure becomes resolved. Such sources are undesirable for a comparison of this kind, since their centroids are not expected to coincide between RACS-Mid1, RACS-High1, and RFC, introducing additional scatter unrelated to the underlying astrometric performance of the catalogues. To ensure a fair comparison between the two epochs, we therefore restrict the RFC analysis to sources that survive the filtering process in both RACS-Mid1 and RACS-High1. In practice, this means considering only RFC sources that are successfully crossmatched with both corrected RACS catalogues. This requirement reduces the final sample to just under 6,000 crossmatches, but ensures that the analysis is performed on a common population of compact, isolated, and bright radio sources suitable for high-precision astrometric comparison.

This aggressive filtering largely removes the influence of source structure, allowing the remaining residuals to approach the limit set by measurement uncertainties and residual catalogue astrometric errors. Consequently, the cores of the offset distributions for the FIRST and RFC comparisons show good agreement, as reflected by their similar 68\% confidence intervals in Table~\ref{tab:racsmid_firstrfc}. The RFC distribution nevertheless exhibits broader tails, particularly in the 95.4\% and 99.7\% confidence intervals, consistent with a small number of remaining sources whose arcsecond-scale ASKAP centroids do not perfectly coincide with their milliarcsecond-scale RFC positions. The confidence intervals, together with the other crossmatching results below a declination of $+30^\circ$, are summarised in Table~\ref{tab:racsmid_firstrfc}.


\begin{table*}[h!]
\centering
\caption{The residual offsets for RACS-High1 vs WISE for scans below Dec. $+30^\circ$.}
\begin{tabular}{|c|cc|cc|cc|}
\hline
\textbf{RACS-High1--WISE Offsets}          & \multicolumn{2}{c|}{\textbf{All-Sky}}                & \multicolumn{2}{c|}{\textbf{On-Plane}}               & \multicolumn{2}{c|}{\textbf{Off-Plane}}              \\ \cline{2-7} 
\textbf{(Dec < $+30^\circ$, arcsec)} & \multicolumn{1}{c|}{\textbf{RA}}    & \textbf{DEC}   & \multicolumn{1}{c|}{\textbf{RA}}    & \textbf{DEC}   & \multicolumn{1}{c|}{\textbf{RA}}    & \textbf{DEC}   \\ \hline
Median                               & \multicolumn{1}{c|}{0.00}           & 0.00           & \multicolumn{1}{c|}{0.00}           & 0.00           & \multicolumn{1}{c|}{0.00}           & 0.00           \\ \hline
68\% Confidence Intervals            & \multicolumn{1}{c|}{-0.17 --- 0.17} & -0.17 --- 0.17 & \multicolumn{1}{c|}{-0.26 --- 0.27} & -0.26 --- 0.25 & \multicolumn{1}{c|}{-0.16 --- 0.16} & -0.17 --- 0.17 \\ \hline
95.4\% Confidence Intervals          & \multicolumn{1}{c|}{-0.44 --- 0.45} & -0.44 --- 0.43 & \multicolumn{1}{c|}{-0.63 --- 0.65} & -0.63 --- 0.62 & \multicolumn{1}{c|}{-0.40 --- 0.40} & -0.40 --- 0.40 \\ \hline
99.7\% Confidence Intervals          & \multicolumn{1}{c|}{-0.92 --- 0.91} & -0.91 --- 0.89 & \multicolumn{1}{c|}{-1.42 --- 1.20} & -1.41 --- 1.31 & \multicolumn{1}{c|}{-0.78 --- 0.77} & -0.80 --- 0.78 \\ \hline
\end{tabular}
\label{tab:racshigh_specs}
\end{table*}

\begin{figure*}[h!]
    \centering
    \includegraphics[scale=0.65]{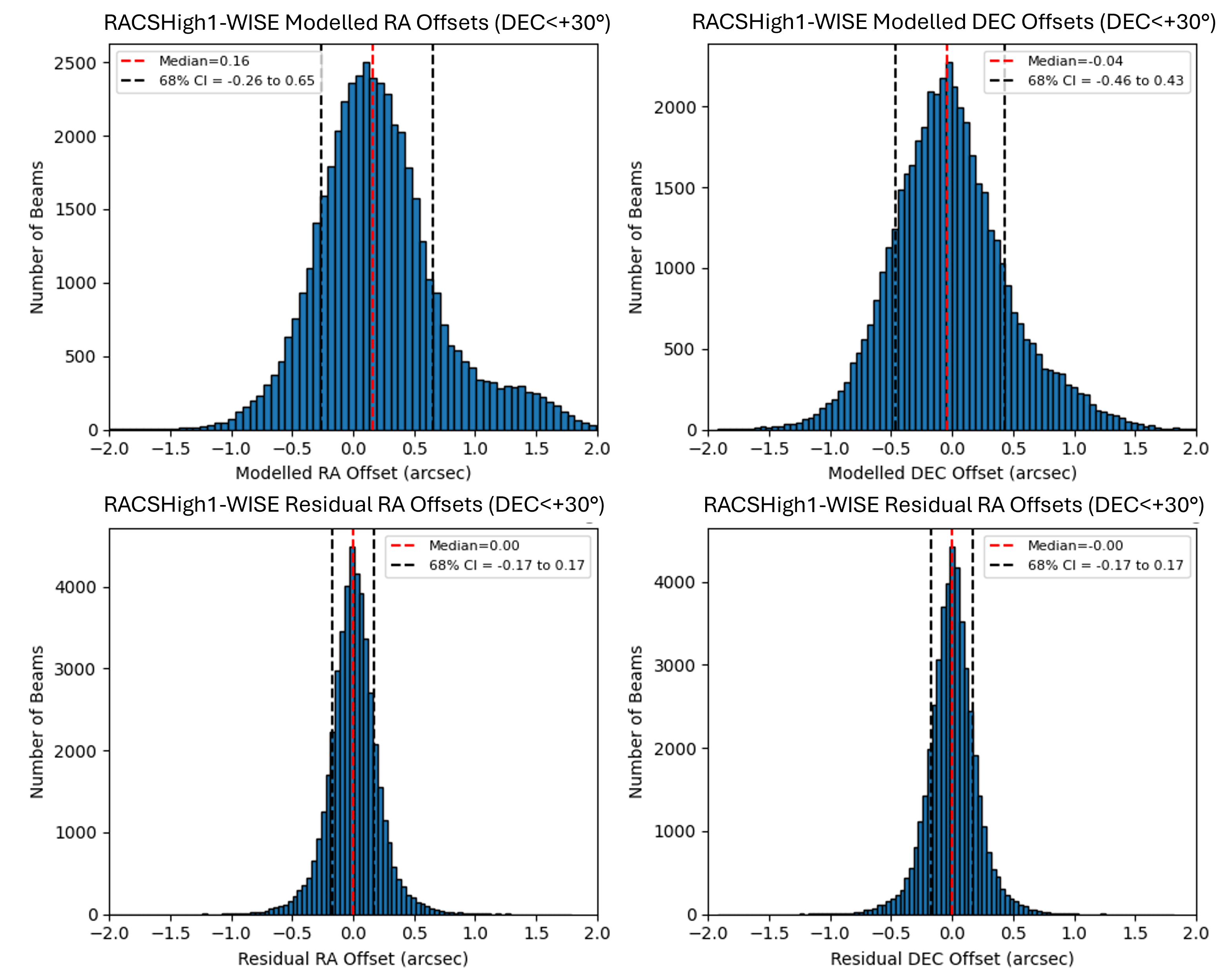}
    \caption{The modelled (top row) and residual (bottom row) offsets for RACS-High1 vs WISE for scans below Dec. $+30^\circ$. The median RA offset (left) improves from $0.16''$ to $0.00''$ and the median Dec. offset (right) decreases from $-0.04''$ to $0.00''$. The 68\% confidence intervals also narrow down significantly.}
    \label{fig:racshigh1histograms}
\end{figure*}

\subsection{RACS-High1 Corrections} \label{sec:racshigh1corrections}

The modelled and residual positional offsets in RACS-High1 relative to WISE are presented in Aitoff projection in Figure~\ref{fig:racshigh1aitoff}. The refined workflow delivers improved corrections across most of the sky, though, as with the other epochs, performance degrades for Dec.\ above $+30^\circ$, where ASKAP operates at very low elevations. We therefore first quantify results for Dec.\ below $+30^\circ$. 

\begin{figure}[h!]
    \centering
    \includegraphics[scale=0.65]{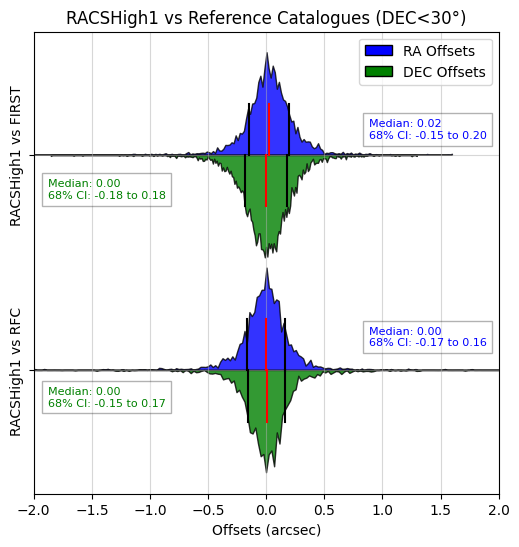}
    \caption{The corrected offsets in RA and Dec. for Dec. < $+30^\circ$ for RACS-High1 vs reference catalogues, with FIRST comparisons in the top row and RFC in the bottom row.}
    \label{fig:racshigh1firstrfc}
\end{figure}

\begin{table*}[h!]
\centering
\caption{Corrected RACS-Mid1 comparison with FIRST and RFC catalogues for regions below Dec. $+30^\circ$}
\begin{tabular}{|c|cc|cc|cc|cc|}
\hline
\textbf{RACS-Mid1 Offsets}     & \multicolumn{2}{c|}{\textbf{vs FIRST All-Sky}}       & \multicolumn{2}{c|}{\textbf{vs RFC All-Sky}}         & \multicolumn{2}{c|}{\textbf{vs RFC On-Plane}}        & \multicolumn{2}{c|}{\textbf{vs RFC Off-Plane}}       \\ \cline{2-9} 
\textbf{(Dec < $+30^\circ$, arcsec)} & \multicolumn{1}{c|}{\textbf{RA}}    & \textbf{DEC}   & \multicolumn{1}{c|}{\textbf{RA}}    & \textbf{DEC}   & \multicolumn{1}{c|}{\textbf{RA}}    & \textbf{DEC}   & \multicolumn{1}{c|}{\textbf{RA}}    & \textbf{DEC}   \\ \hline
Median                               & \multicolumn{1}{c|}{0.03}           & 0.00           & \multicolumn{1}{c|}{0.01}           & 0.01           & \multicolumn{1}{c|}{-0.01}          & -0.00           & \multicolumn{1}{c|}{0.01}           & 0.01           \\ \hline
68\% Confidence Intervals            & \multicolumn{1}{c|}{-0.12 --- 0.17} & -0.16 --- 0.16 & \multicolumn{1}{c|}{-0.17 --- 0.18} & -0.16 --- 0.19 & \multicolumn{1}{c|}{-0.24 --- 0.20} & -0.18 --- 0.23 & \multicolumn{1}{c|}{-0.16 --- 0.18} & -0.15 --- 0.19 \\ \hline
95.4\% Confidence Intervals          & \multicolumn{1}{c|}{-0.32 --- 0.37} & -0.37 --- 0.38 & \multicolumn{1}{c|}{-0.52 --- 0.51} & -0.51 --- 0.61 & \multicolumn{1}{c|}{-0.67 --- 0.46} & -0.52 --- 0.72 & \multicolumn{1}{c|}{-0.51 --- 0.52} & -0.51 --- 0.59 \\ \hline
99.7\% Confidence Intervals          & \multicolumn{1}{c|}{-0.80 --- 0.76} & -0.80 --- 0.76 & \multicolumn{1}{c|}{-1.80 --- 1.43} & -1.84 --- 2.00 & \multicolumn{1}{c|}{-2.07 --- 1.31} & -2.96 --- 1.92 & \multicolumn{1}{c|}{-1.51 --- 1.43} & -1.60 --- 2.00 \\ \hline
\end{tabular}
\label{tab:racsmid_firstrfc}
\end{table*}

\begin{table*}[h!]
\centering
\caption{Corrected RACS-High1 comparison with FIRST and RFC catalogues for regions below Dec. $+30^\circ$}
\begin{tabular}{|c|cc|cc|cc|cc|}
\hline
\textbf{RACS-High1 Offsets}          & \multicolumn{2}{c|}{\textbf{vs FIRST All-Sky}}       & \multicolumn{2}{c|}{\textbf{vs RFC All-Sky}}         & \multicolumn{2}{c|}{\textbf{vs RFC On-Plane}}        & \multicolumn{2}{c|}{\textbf{vs RFC Off-Plane}}       \\ \cline{2-9} 
\textbf{(Dec < $+30^\circ$, arcsec)} & \multicolumn{1}{c|}{\textbf{RA}}    & \textbf{DEC}   & \multicolumn{1}{c|}{\textbf{RA}}    & \textbf{DEC}   & \multicolumn{1}{c|}{\textbf{RA}}    & \textbf{DEC}   & \multicolumn{1}{c|}{\textbf{RA}}    & \textbf{DEC}   \\ \hline
Median                               & \multicolumn{1}{c|}{0.02}           & 0.00           & \multicolumn{1}{c|}{0.00}           & 0.00           & \multicolumn{1}{c|}{-0.02}          & 0.00           & \multicolumn{1}{c|}{0.00}          & 0.00           \\ \hline
68\% Confidence Intervals            & \multicolumn{1}{c|}{-0.15 --- 0.20} & -0.18 --- 0.18 & \multicolumn{1}{c|}{-0.17 --- 0.16} & -0.15 --- 0.17 & \multicolumn{1}{c|}{-0.23 --- 0.15} & -0.17 --- 0.20 & \multicolumn{1}{c|}{-0.16 --- 0.16} & -0.15 --- 0.16 \\ \hline
95.4\% Confidence Intervals          & \multicolumn{1}{c|}{-0.42 --- 0.46} & -0.46 --- 0.44 & \multicolumn{1}{c|}{-0.52 --- 0.45} & -0.46 --- 0.54 & \multicolumn{1}{c|}{-0.66 --- 0.46} & -0.50 --- 0.64 & \multicolumn{1}{c|}{-0.49 --- 0.45} & -0.44 --- 0.52 \\ \hline
99.7\% Confidence Intervals          & \multicolumn{1}{c|}{-0.89 --- 0.94} & -0.98 --- 0.85 & \multicolumn{1}{c|}{-1.44 --- 1.31} & -1.52 --- 1.73 & \multicolumn{1}{c|}{-1.99 --- 1.26} & -2.84 --- 2.80 & \multicolumn{1}{c|}{-1.36 --- 1.31} & -1.39 --- 1.64 \\ \hline
\end{tabular}
\label{tab:racshigh_firstrfc}
\end{table*}

The histograms shown in Figure~\ref{fig:racshigh1histograms} demonstrate that the mean bias in both RA and Dec.\ is effectively removed, while the overall scatter is substantially reduced. A summary of the resulting statistics is provided in Table~\ref{tab:racshigh_specs}. Again, the residual error distributions deviate from a purely Gaussian form, exhibiting extended tails that are captured quantitatively in these results. This behaviour becomes more pronounced within the Galactic plane, where both the scatter and the degree of non-Gaussianity increase, reflecting the reduced effectiveness of the corrections in these more complex regions.



When considering the full RACS-High1 sky, including regions above Dec.\ $+30^\circ$, the mean bias remains eliminated and the scatter is again reduced, with the 68\% confidence intervals improving from $0.11''^{+0.60}_{-0.39}$ to $0.00''^{+0.18}_{-0.18}$ in RA and from $0.01''^{+0.68}_{-0.43}$ to $0.00''^{+0.19}_{-0.19}$ in Dec. As expected, some degradation persists in the northern declinations and in the Galactic plane, but the corrections remain robust across the majority of the survey coverage. It is worth noting that the corrections for the high-frequency epoch are marginally less precise than those for the mid-frequency epoch. This is unlikely to be driven by ionospheric effects, which generally diminish at higher observing frequencies under comparable conditions. Instead, the slightly larger residual scatter is more plausibly attributed to the reduced source density and lower signal-to-noise ratio at higher frequencies, which limit the number of compact reference sources available for robust averaging within each beam. Additionally, the RACS-High1 observations were conducted later in time and may have coincided with periods of elevated solar activity, potentially introducing additional calibration variability in some regions.

\begin{figure*}[h!]
    \centering
    \includegraphics[scale=0.65]{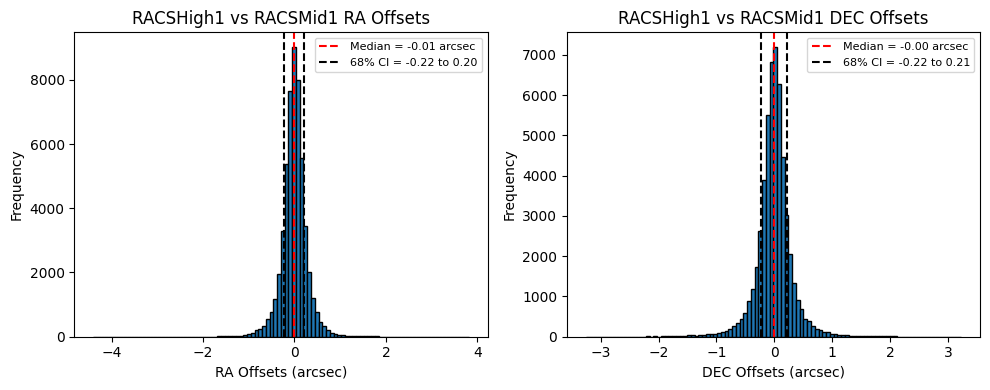}
    \caption{The corrected RACS-High1 source lists compared to the corrected RACS-Mid1 source lists in RA and Dec. The results show that, post corrections, both the source lists are very comparable to each other.}
    \label{fig:racshigh1_mid1}
\end{figure*}

External validation was again carried out using FIRST and RFC (Figure~\ref{fig:racshigh1firstrfc}), again following the procedure outlined in \citet{jaini2025enhanced}, with some modifications already described in Section~\ref{sec:racsmid1corrections}. As in the mid-band analysis, VLASS was not used for verification since it contributed to the Stage~1 corrections.  


Crossmatching with FIRST yielded median post-correction offsets of $0.01''$ in RA and $0.00''$ in Dec., based on $\sim$9,000 overlapping beams with an average of 40 crossmatched sources per beam. Again, the RFC comparison is restricted to sources that satisfy the filtering criteria in both RACS-Mid1 and RACS-High1, ensuring that the analysis is performed on a common population of compact, isolated, and bright radio sources. When these filtered RACS-High1 sources are crossmatched individually with RFC using a matching radius of $5''$, we obtain median offsets of $0.00''$ in both RA and Dec.\ for just under 6,000 crossmatches. The confidence intervals, together with the remaining crossmatching statistics below a declination of $+30^\circ$, are summarised in Table~\ref{tab:racshigh_firstrfc}.

As with RACS-Mid1, the cores of the FIRST and RFC offset distributions show good agreement, while the RFC distribution exhibits broader tails due to a small number of remaining sources whose arcsecond-scale ASKAP centroids do not perfectly coincide with their milliarcsecond-scale RFC positions. Overall, these results confirm that the RACS-High1 corrections achieve a level of astrometric precision comparable to that obtained for RACS-Mid1, and demonstrate that the correction methodology remains robust across the full ASKAP frequency range.


\begin{figure}
    \centering
    \includegraphics[width=\textwidth]{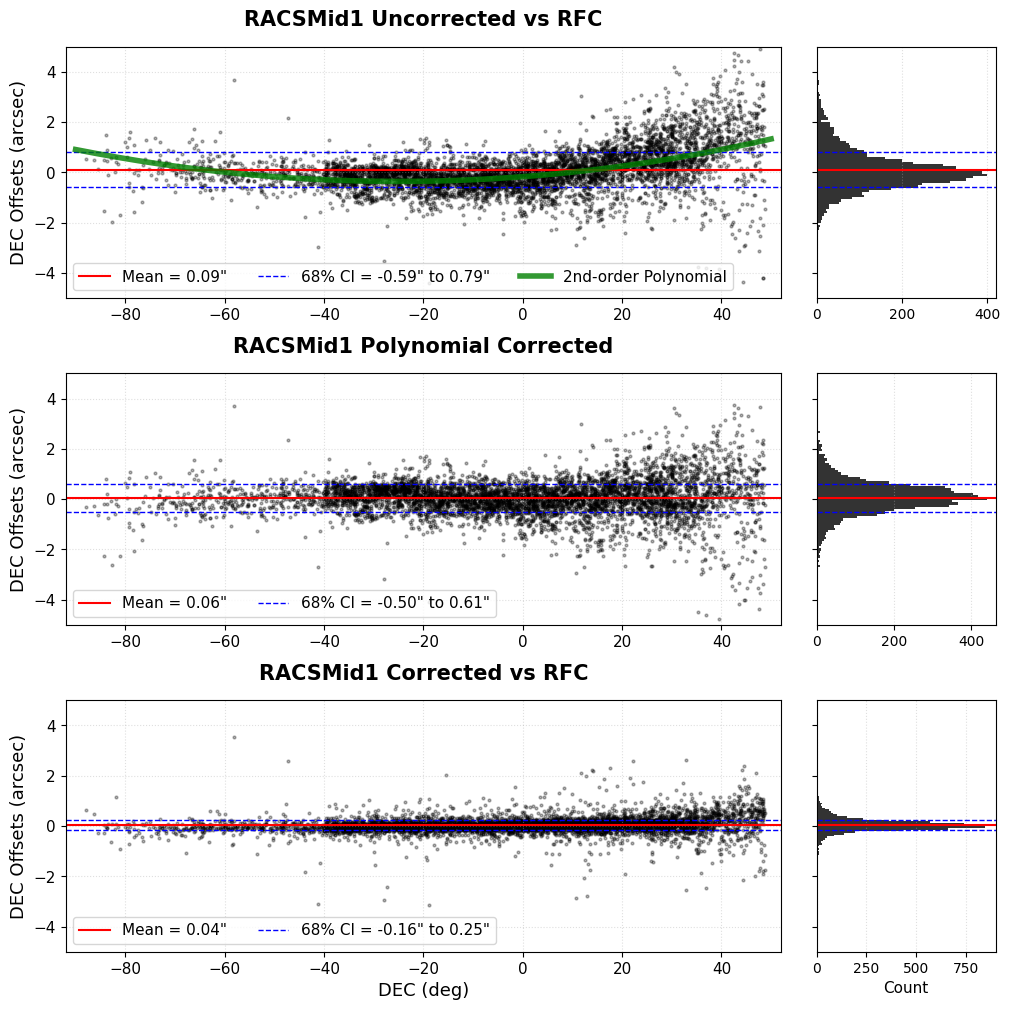}
    \caption{RACS-Mid1 astrometric offsets in Dec., for $\sim6,000$ individual RFC sources, plotted as a function of Dec. The top panel shows the uncorrected offsets having a mean of $0.09''$ and 68\% confidence intervals between $[+0.79'', -0.59'']$, exhibiting systematic declination-dependent errors. The panel also shows a 2nd-order polynomial (in green) fitting the scatter, as described in \citet{2024PASA...41....3D}. The middle panel shows the offsets after subtracting the polynomial to have a mean of $0.06''$ and 68\% confidence intervals of $[+0.61'', -0.50'']$, significantly reducing the largest outliers, but still exhibiting considerable scatter. The bottom panel shows the offsets after applying our corrections to have a mean of $0.04''$ and 68\% confidence intervals ranging between $[+0.25'', -0.16'']$, illustrating the both the removal of the systematic trend and a much greater reduction in scatter. The reader should note that these values are computed over the full declination range of RACS-Mid1, whereas the results presented in Table~\ref{tab:racsmid_firstrfc} are restricted to declinations below $+30^\circ$. As a result, the quoted values differ slightly. The right-hand panels show the collapsed histograms of the offsets.}
    \label{fig:racsmid_rfc_full}
\end{figure}

\begin{figure}
    \centering
    \includegraphics[width=\textwidth]{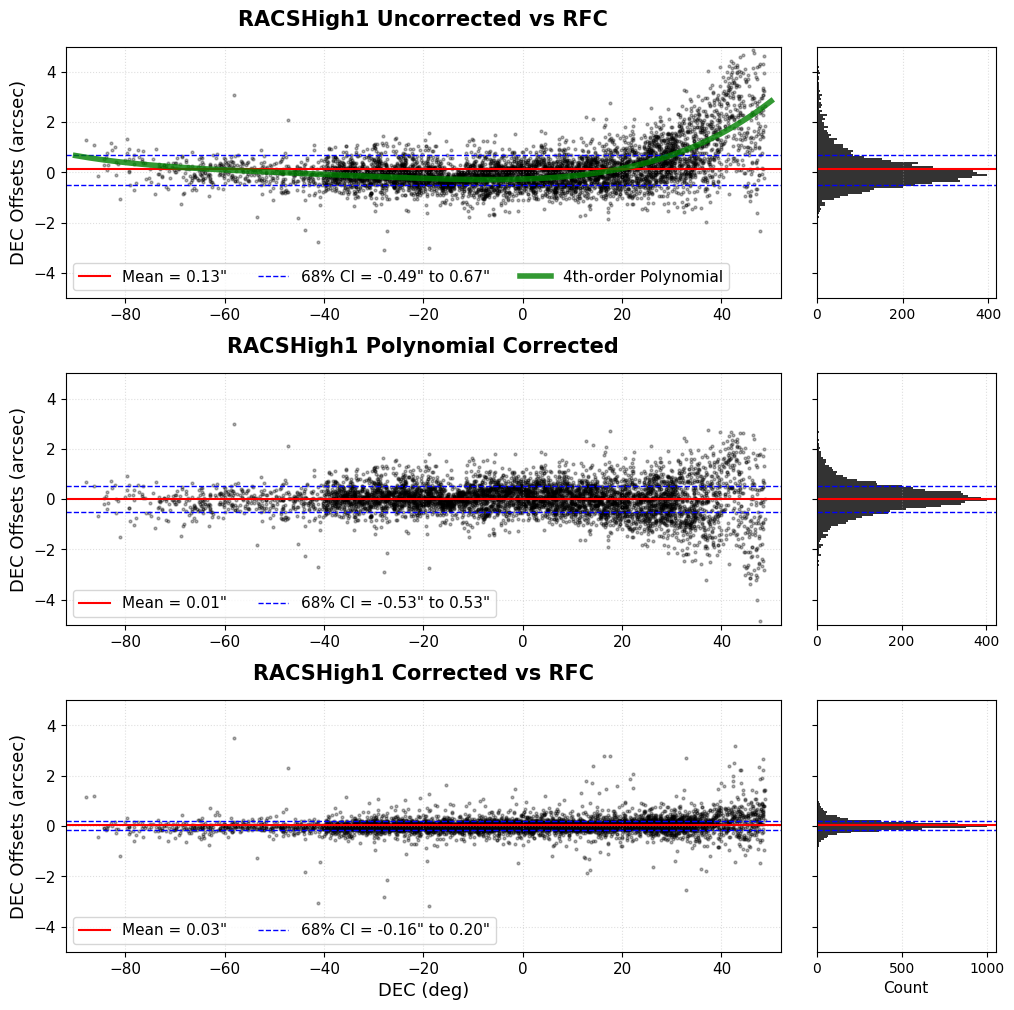}
    \caption{RACS-High1 astrometric offsets in Dec., for $\sim6,000$ individual RFC sources, plotted as a function of Dec., following the same structure as Figure~\ref{fig:racsmid_rfc_full}. The uncorrected data (top) shows significant scatter having a mean of $0.13''$ and 68\% confidence intervals ranging between [$+0.67'', -0.49'']$, along with a 4th-order polynomial (in green) fitting the scatter, as described in \citet{duchesne2025rapid}. The polynomial corrected plot (middle) shows slightly reduced scatter with a mean of $0.01''$ and 68\% confidence intervals of $[+0.53'', -0.53'']$. The offsets are markedly reduced in the data corrected using our methodology (bottom) to a mean of $0.03''$ and 68\% confidence intervals between $[+0.20'', -0.16'']$. The reader should note that these values are computed over the full declination range of RACS-High1, whereas the results presented in Table~\ref{tab:racshigh_firstrfc} are restricted to declinations below $+30^\circ$. As a result, the quoted values differ slightly. The right-hand panels show the collapsed histograms of the offsets.}
    \label{fig:racshigh_rfc_full}
\end{figure}

\subsection{Cross-Epoch Comparisons} \label{sec:racsmid1high1low1}

To assess consistency across the corrected catalogues, we performed a direct crossmatch of corrected RACS-Mid1 and RACS-High1 sources with each other. Median offsets between catalogues are all within $\pm0.01''$, with 68\% confidence intervals consistent within the noted uncertainties. Histograms of these comparisons are shown in Figure~\ref{fig:racshigh1_mid1}. Further comparison with RACS-Low3 also showed similar results for both RACS-Mid1 and RACS-High1. For RA, the median offsets were within $\pm0.1''$ and 68\% confidence intervals consistent within $\pm0.3''$. For Dec., the median offsets were within $\pm0.09''$ and 68\% confidence intervals consistent within $\pm0.4''$. This degradation in results is due to the slightly worse corrections of RACS-Low3 compared to the rest of the epochs, and can also be seen in the RACS-Low3 results published in \citet{jaini2025enhanced}.

While minor discrepancies remain in fields with low crossmatch density or elevated source confusion, particularly within the Galactic plane, the corrections are robust across the vast majority of the survey footprint. These validations support the use of the corrected catalogues for precise sky localisation in current and future multi-wavelength studies.

\bigskip

It is important to note that the methods used to estimate the residual astrometric errors each have inherent limitations. Comparisons with FIRST are restricted in sky coverage and involve averaging over beam-scale regions, which can lead to slightly optimistic estimates of the residual errors. In contrast, comparisons with the RFC catalogue are based on individual source matches and can be affected by intrinsic differences between milliarcsecond-scale VLBI positions and arcsecond-scale ASKAP centroids, although this effect is mitigated as far as possible through source filtering. Nevertheless, combining these two estimates of the astrometric residuals, we see evidence that the residual error distribution is not perfectly modelled by a Gaussian distribution, and both very small and very large residuals arise more frequently than would be expected from a Gaussian distribution.


Given this non-Gaussian behaviour, we recommend that for applications requiring high-precision astrometry, users consult the sky maps of residual offsets to assess local variations in accuracy. However, to provide a practical and broadly applicable estimate, we adopt 1-$\sigma$ systematic uncertainties of $0.25''$ outside the Galactic plane and $0.35''$ within the plane. These values are based on the measured standard deviations of the residual offset distributions in the RFC comparisons rounded down to two decimal places. These values are also consistent with, or slightly conservative relative to, the 95\% confidence intervals reported in Tables~\ref{tab:racsmid_firstrfc} and~\ref{tab:racshigh_firstrfc}, while remaining appropriately cautious given the extended tails observed in the residual distributions.



\section{DISCUSSION} \label{sec:discussion}

\subsection{Comparison with Polynomial Corrections}

The public data releases of RACS-Mid1 (\citealp{2024PASA...41....3D}) and RACS-High1 (\citealp{duchesne2025rapid}) identified a systematic declination-dependent astrometric offset in the uncorrected catalogues. In those studies, the dominant trend was characterised as a function of declination alone and corrected using global polynomial fits relative to the ICRF3 (\citealp{2020A&A...644A.159C}) reference frame. Specifically, a second-order polynomial was applied to the RACS-Mid1 catalogue and a fourth-order polynomial to the RACS-High1 catalogue. While these models successfully recentred the mean offsets, the residual scatter remained substantial with offsets reaching up to $\pm2''$ in RACS-Mid1 and $\pm2.5''$ in RACS-High1 prior to correction.

In contrast, our approach operates at the beam and scan level, modelling calibration errors as discrete instrumental effects rather than as a smooth all-sky distortion. By working directly with unmosaicked per-beam source lists and solving for scan-independent and beam-independent offset terms, we capture the physical structure of the astrometric errors more accurately. To directly compare these two correction strategies, we evaluated the performance of our solution relative to the polynomial-corrected datasets.

For this comparison we make use of the corrected RACS-Mid1 and RACS-High1 source lists with the same aggressive filtering described in Section~\ref{sec:racsmid1corrections}. As a reference, we use the RFC catalogue with a crossmatching radius of $5''$, which provides a denser grid of compact, high-precision radio reference sources and samples the sky more finely than ICRF3. Because the polynomial corrections depend only on declination, the RFC comparison allows us to examine the resulting astrometric residuals across the full declination range without introducing additional spatial averaging. Figures~\ref{fig:racsmid_rfc_full} and~\ref{fig:racshigh_rfc_full} show the positional offsets in Dec.\ as a function of declination for RACS-Mid1 and RACS-High1, respectively.

In the uncorrected data (top panels), both epochs exhibit a clear declination-dependent systematic trend. After applying our corrections (bottom panels), this trend is effectively removed across the entire declination range, with the residual offsets tightly centred around zero. Compared with the polynomial-based corrections, our beam-resolved modelling reduces the residual scatter by approximately a factor of two. This comparison indicates that the dominant astrometric systematics in RACS arise from local calibration behaviour, and are therefore more effectively corrected through beam-resolved modelling than through a purely global fit.

\begin{figure*}[h!]
    \centering
    \includegraphics[scale=0.60]{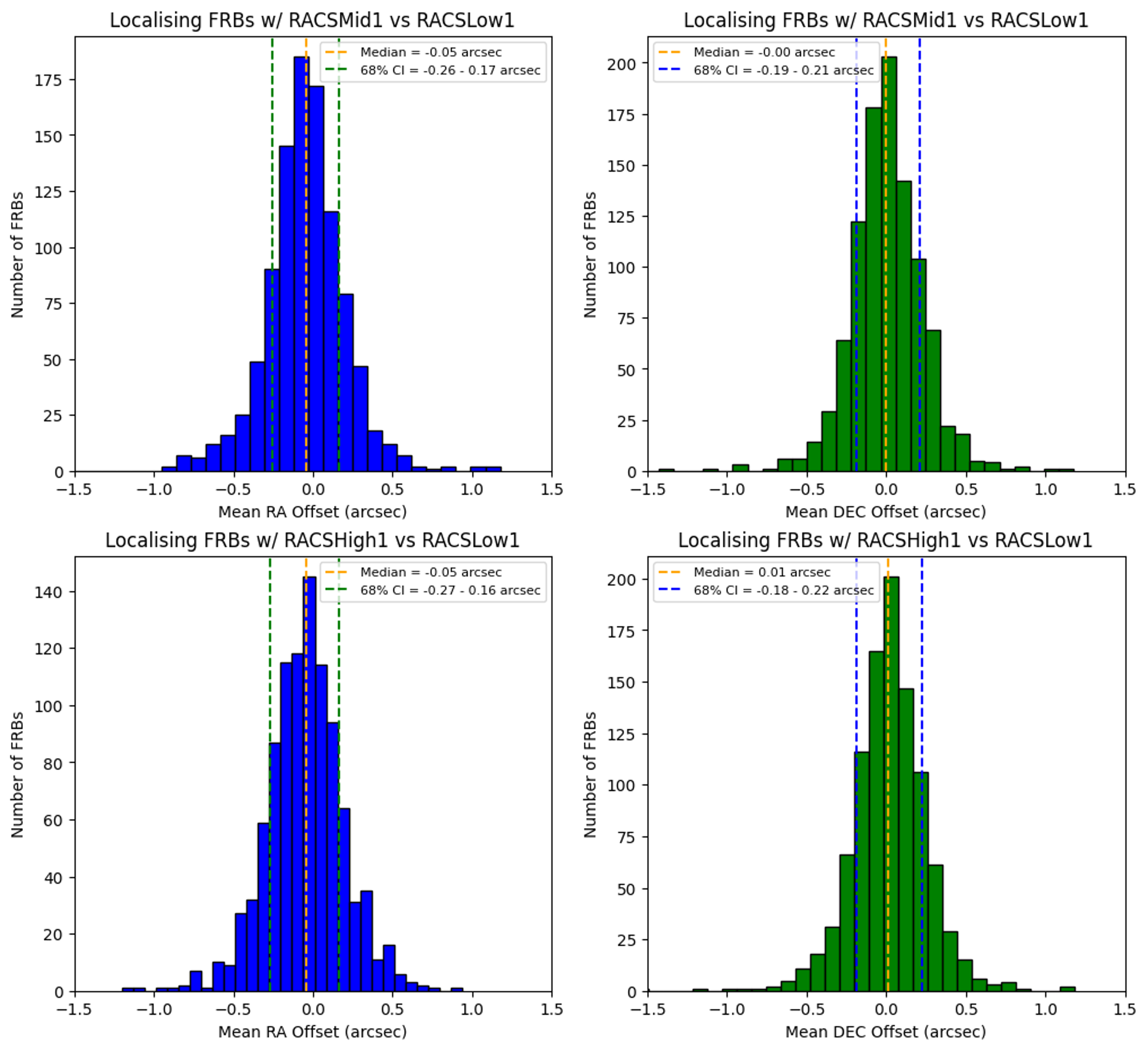}
    \caption{The ensemble mean localisation offsets for 1000 simulated FRBs randomly distributed across the ASKAP sky ($-80^\circ < \mathrm{Dec.} < +25^\circ$). The offsets obtained using the corrected RACS-Mid1 (top) and RACS-High1 (bottom) source lists are shown relative to those derived using the corrected RACS-Low1 catalogue. The left column shows the differences in RA, while the right column shows the corresponding differences in Dec.}
    \label{fig:racshigh1_mid1_low1}
\end{figure*}

\subsection{Modelling the Use of RACS in ASKAP FRB Localisations}

As described in Section~\ref{sec:introduction}, one of the primary motivations for this work was to maximise the precision of ASKAP FRB localisation. Having now implemented our corrections on all three RACS bands, we now simulate the effect of using these as reference catalogues for FRB localisations. The analysis is designed to emulate the behaviour of the CRAFT (the Commensal Realtime ASKAP Fast Transients survey, \citealp{2010PASA...27..272M}) Effortless Localisation and Enhanced Burst Inspection (CELEBI, \citealp{2023A&C....4400724S}) pipeline. Previously, mid- and high-band ASKAP images relied on RACS-Low1 as the astrometric reference. While globally consistent at the $\sim0.25''$ level, such cross-frequency referencing can introduce small but non-negligible systematic shifts. The aim of this section is therefore threefold: (i) to assess the impact of residual direction-dependent astrometric errors between epochs, (ii) to evaluate the role of intrinsic source-structure differences with frequency and angular resolution, and (iii) to examine how ensemble-selection effects influence the mean offset used in FRB localisation.

We generated 1000 random FRB positions uniformly distributed across the RACS survey footprint between Dec.\ $-80^\circ$ and $+25^\circ$. For each simulated field, we queried sources within a $1^\circ$ radius from the corrected RACS-Low1 Gaussian component catalogue and from the corrected RACS-Mid1 and RACS-High1 source lists. This mirrors the behaviour of the CELEBI localisation pipeline (Section~\ref{sec:presentuse}) and therefore captures the practical astrometric behaviour relevant to ASKAP transient follow-up.


ASKAP FRB detections trigger the download of $\sim12$~s of raw voltage data (\citealp{james2025nanosecond}), which are imaged to produce a field with a typical root-mean-square noise of a few mJy (\citealp{2023A&C....4400724S}). In practice, only sources detected at $\gtrsim5$--$6\sigma$ contribute to localisation measurements, corresponding to peak flux densities of $\gtrsim15$~mJy in these short integrations. To emulate the population of continuum sources that would realistically be detectable in such images, we first restricted our simulated catalogue queries to sources above this effective detection threshold. We then applied additional filtering steps designed to mimic the source-selection strategy used in the CELEBI pipeline (for more details, refer to Section~3.7 of \citealp{2023A&C....4400724S}). These steps remove sources that are resolved, blended, or morphologically complex, since their centroid positions may depend on the observing frequency or synthesised beam. Because the beam shape and angular resolution of the FRB follow-up image generally differ from those of the RACS observations, such sources can exhibit frequency- or resolution-dependent centroid shifts that would bias the measured field offset. The filtering heuristic therefore assigns each source an effective flux density that accounts for the brightness contrast with its nearest neighbour and imposes a brightness-dependent isolation criterion, ensuring that the ensemble used for offset estimation is dominated by compact, isolated sources whose centroids are stable across frequencies and angular resolutions.

Let $S$ and $S_n$ denote the peak flux densities of a source and its nearest neighbour. We define the flux ratio

\begin{equation}
    r = \frac{S}{S_n + \varepsilon},
\end{equation}
where $\varepsilon$ is a small positive constant of $10^{-6}$ preventing division by zero.

An adjusted (or effective) flux density is then computed as:

\begin{equation}
    S_{\mathrm{eff}} = S \sqrt{\max(r,\varepsilon)},
\end{equation}
which increases when the primary source is significantly brighter than its neighbour, and decreases when the neighbour has comparable or higher brightness. 

The required minimum angular separation is derived through a one-dimensional linear interpolation:

\begin{equation}
    \mathrm{\theta_{min}} = \mathrm{interp}\!\left(S_{\mathrm{eff}}; \{S_i\} \rightarrow \{\Delta\theta_i\}\right),
\end{equation}
where the flux nodes $\{S_{i}\}$ (e.g., 25, 50, 100~mJy) are mapped to corresponding angular separations $\{\Delta \theta_{i}\}$ (e.g., 10, 20, 30~arcsec).

A source passes the isolation criterion if its measured nearest-neighbour separation $\Delta \theta$ satisfies:

\begin{equation}
    \Delta\theta > \mathrm{\theta_{min}}.
\end{equation}
This isolation mask is applied in combination with existing compactness and SNR-based cuts, forming the final selection of isolated, point-like sources used for crossmatching, mimicking the strategy used in the CELEBI pipeline.

This method is intentionally conservative and numerically robust: the $\sqrt{r}$ scaling moderates extreme flux ratios so that exceptionally bright contrasts do not impose unrealistic separation requirements, while the clipping and inclusion of a small $\varepsilon$ prevent numerical instabilities. Faint isolated sources remain usable, whereas dominant bright sources must satisfy stricter separation criteria, reducing bias from blends. The tuning parameters were optimised through visual inspection and validated via injection–recovery tests. These parameters may be adjusted in future work to trade completeness against purity, depending on the scientific goals of the analysis. A variation of this heuristic has also been integrated into the CELEBI pipeline (\citealp{celebi2.0}).

After filtering, we crossmatched the RACS-Mid1 and RACS-High1 source lists with RACS-Low1 to identify common sources in each simulated FRB field and computed the mean RA and Dec.\ offsets as shown in Figure~\ref{fig:racshigh1_mid1_low1}. For RACS-Mid1, we identified an average of 12.5 common sources per field, yielding median offsets of $-0.05''^{+0.17}_{-0.26}$ in RA and $0.00''^{+0.21}_{-0.19}$ in Dec. For RACS-High1, the average decreased to 8.1 common sources per field, with median offsets of $-0.05''^{+0.16}_{-0.27}$ in RA and $0.01''^{+0.22}_{-0.18}$ in Dec. These statistics confirm that all three corrected epochs remain mutually consistent at the $\sim 0.25''$ level over most of the sky.

\begin{figure*}[h!]
    \centering
    \includegraphics[width=\textwidth]{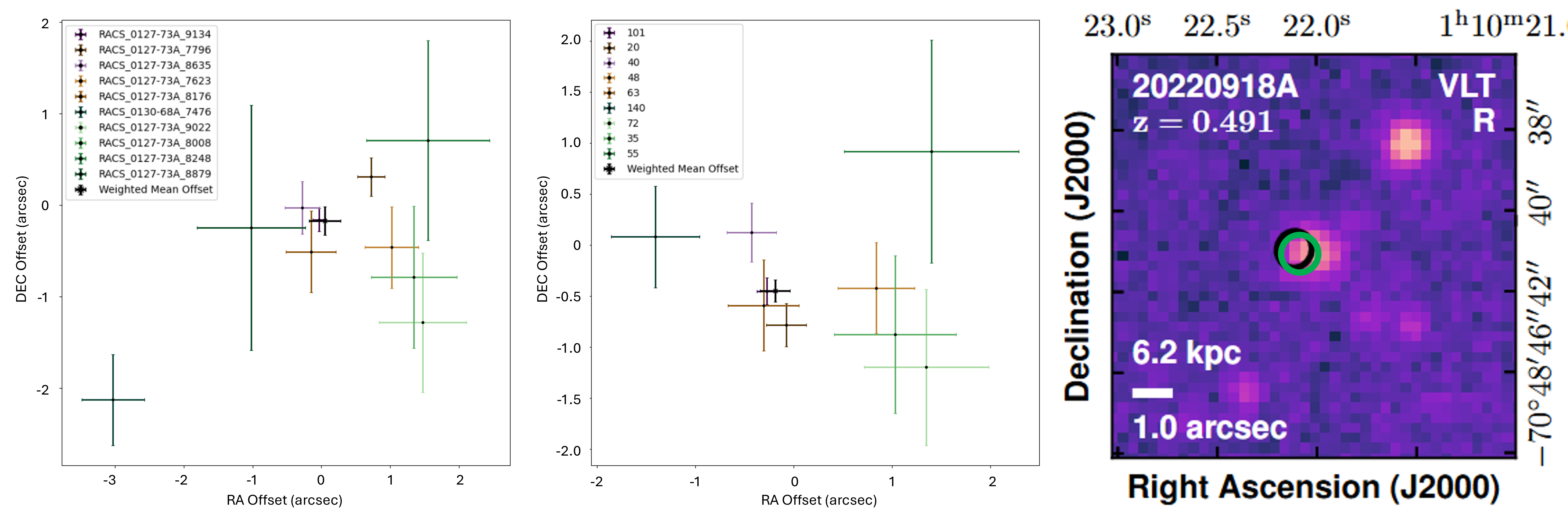}
    \caption{An example of determining the positional offsets of the mid-band FRB20220918A while using the corrected RACS-Low1 catalogues (left) and the corrected RACS-Mid1 source lists (middle) with CELEBI. After updating the RACS epoch, the estimated offset correction based on the field sources changed from $0.05'' \pm 0.23''$, $-0.17'' \pm 0.16''$ to $-0.19'' \pm 0.15''$, $-0.45'' \pm 0.11''$ in RA and Dec. respectively. After including an assumed 1-$\sigma$ systematic uncertainty of $0.3''$ (as the source is outside the Galactic plane) and the statistical FRB position uncertainty, the final position of the FRB is updated from $1^{h}10^{m}22^{s}.10 \pm 0.35''$ to $1^{h}10^{m}22^{s}.06 \pm 0.35''$ in RA and from $-70^\circ48'40''.94 \pm 0.37''$ to $-70^\circ48'41''.21 \pm 0.37''$ in Dec. The effect of this change on the FRB's localisation within the host galaxy is shown in the VLT R-band image of the host (right), where the black error ellipse indicates the localisation with the corrected RACS-Low1 catalogues (\citealp{shannon2025commensal}), and the green ellipse shows the localisation achieved with the corrected RACS-Mid1 source lists.}
    \label{fig:celebi_220918}
\end{figure*}

\begin{figure*}[h!]
    \centering
    \includegraphics[width=\textwidth]{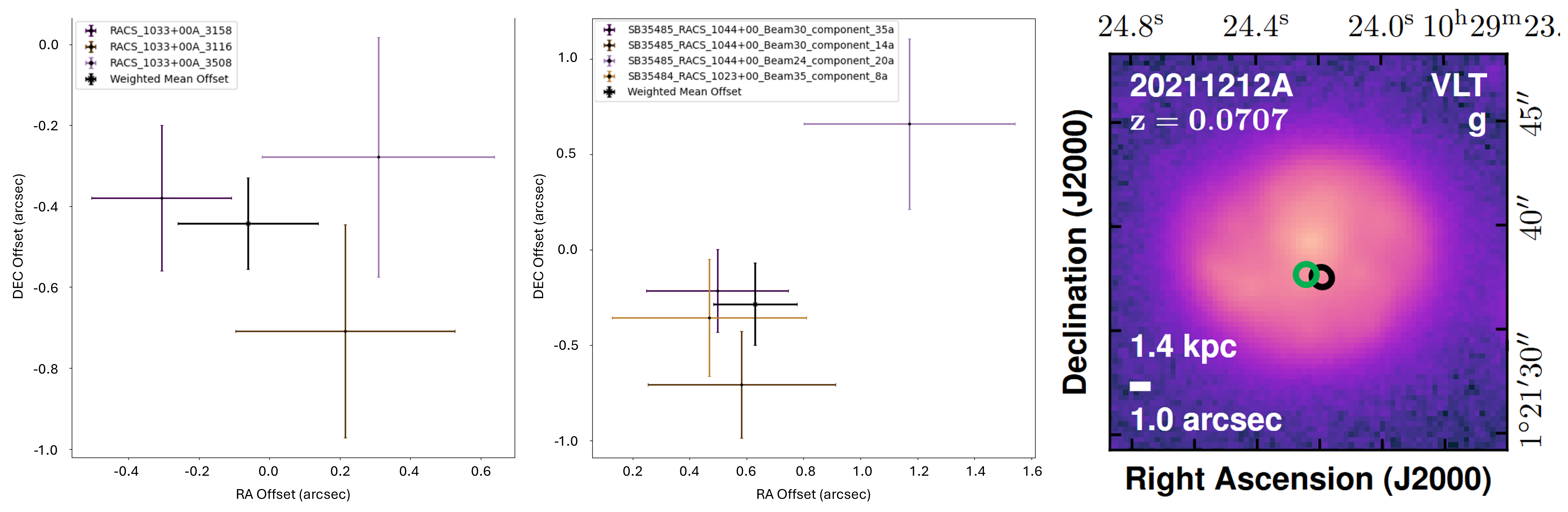}
    \caption{An example of determining the positional offsets of high-band FRB20211212A while using the corrected RACS-Low1 catalogues (left) and the corrected RACS-High1 source lists (middle) with CELEBI. After updating the RACS epoch, the estimated offset correction based on the field sources changed from $-0.06'' \pm 0.20''$, $0.44'' \pm 0.11''$ to $0.63'' \pm 0.15''$, $-0.28'' \pm 0.22''$ in RA and Dec. respectively. A large part of the shift in the offsets can be contributed to the additional source that was crossmatched while using the RACS-High1 source lists. This could be because our heuristic is not as refined with the mid- and high-band source lists when compared to the low-band heuristic, due to the lack of mosaicked source catalogues in addition to the Gaussian components represented by the source lists. After including an assumed 1-$\sigma$ systematic uncertainty of $0.3''$ (as the source is well outside the Galactic plane) and the statistical FRB position uncertainty, the final position of the FRB is updated from $10^{h}29^{m}24^{s}.19 \pm 0.40''$ to $10^{h}29^{m}24^{s}.24 \pm 0.42''$ in RA and from $+01^\circ21'37''.55 \pm 0.40''$ to $+01^\circ21'37''.71 \pm 0.42''$ in Dec. The effect of this change on the FRB's localisation within the host galaxy is shown in the VLT g-band image of the host (right), where the black error ellipse indicates the localisation with the corrected RACS-Low1 catalogues (\citealp{shannon2025commensal}), and the green ellipse shows the localisation achieved with the corrected RACS-High1 source lists.}
    \label{fig:celebi_211212}
\end{figure*}

However, the field-to-field scatter seen in the simulations reveals measurable differences between epochs. These differences can be understood by considering the different contributions to the ensemble mean offset measured for each FRB field. The minimum contribution arises from the residual direction-dependent astrometric errors that remain in the corrected RACS catalogues. As shown in Section~\ref{sec:results}, comparisons with external reference catalogues such as FIRST and RFC indicate that these residual systematic errors are typically at the level of $\sim0.25''$ across most of the sky. This therefore represents the fundamental floor on the achievable localisation precision when using RACS catalogues to determine field offsets.

A second contribution arises from the statistical uncertainty associated with estimating the mean offset from a finite ensemble of reference sources. In practice this term is generally small. Even after applying the filtering criteria described above, each simulated FRB field typically retains $\sim60$ usable sources within a $1^\circ$ radius. The resulting statistical uncertainty on the ensemble mean is therefore well below the residual catalogue astrometric errors and does not significantly broaden the offset distribution.

The remaining contribution is due to frequency- and resolution-dependent centroid shifts in real sources. These arise because the ASKAP synthesised beam and observing frequency may differ from those used to construct the reference catalogue, particularly when a catalogue from a different RACS band is used to estimate the field offset. The effect can be seen by comparing the simulated offsets in Figure~\ref{fig:racshigh1_mid1_low1} with the catalogue crosschecks presented in Figures~\ref{fig:racsmid1firstrfc} and~\ref{fig:racshigh1firstrfc}. When the reference catalogue is drawn from a different observing band, the resulting astrometric precision is slightly worse than the baseline residual errors measured in the catalogue validation tests. This indicates that the dominant contribution to the localisation uncertainty remains the residual direction-dependent calibration errors in RACS, but that an additional, smaller component arises from frequency- and resolution-dependent centroid differences.

Taken together, these results show that the use of a frequency-matched reference catalogue minimises the additional contribution from source-structure effects and therefore approaches the best achievable precision set by the residual RACS astrometric errors. Maintaining well-calibrated catalogues for each RACS observing band therefore improves the robustness of ASKAP transient localisation, allowing the offset estimation to operate as close as possible to the intrinsic astrometric precision of the survey.

\subsection{Present use of the RACS-Mid1 and RACS-High1 Corrections} \label{sec:presentuse}

The corrected RACS-Mid1 and RACS-High1 source lists have now been incorporated into the CELEBI pipeline, alongside the corrected RACS-Low1 catalogues, to refine the positional accuracy of ASKAP-discovered FRBs. Unlike RACS-Low1, which is available as a compiled catalogue, both RACS-Mid1 and RACS-High1 are currently accessible only as unmosaicked per-beam source lists. This structural difference affects how these datasets are integrated into CELEBI, requiring additional handling of the beam-wise source lists during the localisation process. This also means that the heuristic we used for crossmatching with RACS-Mid1 and RACS-High1 is not as sophisticated as the one we developed for RACS-Low1. A review of this expanded functionality and its limitations, together with other improvements to the pipeline, will be presented in \citet{celebi2.0}.

To illustrate the practical benefits of these corrections, we highlight two case studies of FRBs discovered at higher frequencies. 

Figure~\ref{fig:celebi_220918} presents the case of FRB20220918A, an ASKAP mid-band FRB discovered at a central frequency of 1271.5 MHz and a redshift of 0.491. When the corrected RACS-Low1 catalogue is used as the reference, the weighted mean offset in its position differs slightly compared to when the corrected RACS-Mid1 source lists are employed. This positional change, although on the order of the 1-$\sigma$ astrometric uncertainty, is sufficient to alter the probability of associating the FRB with nearby galaxies in deep optical images, especially at such a high redshift. The shift can be visualised in the VLT/FORS2\footnote{\url{eso.org/public/teles-instr/paranal-observatory/vlt/vlt-instr/fors}} (the Very Large Telescope - FOcal Reducer and low dispersion Spectrograph 2) image, where the localisation ellipse moves subtly across the candidate host galaxy, underscoring the importance of band-matched reference catalogues.

Similarly, Figure~\ref{fig:celebi_211212} shows the case of FRB20211212A, the only ASKAP high-band FRB discovered to date, at a central frequency of 1631.5 MHz. For this event, we compare the offsets when the corrected RACS-Low1 catalogue is used as the reference against those obtained using the corrected RACS-High1 source lists. As with the mid-band example, the changes are modest but significant: the position shifts relative to the VLT/FORS2 imaging alter the likelihood of associating the burst with faint background galaxies in the vicinity. In this case, the corrections help reduce the residual systematic bias that arises when a lower-frequency catalogue is used to calibrate higher-frequency FRBs, thereby providing a truer match between the observing band and the reference frame.

For this comparison, we adopt a common 1-$\sigma$ astrometric uncertainty of $0.3''$ for all RACS epochs. Although the RACS-Mid1 and RACS-High1 corrections demonstrate slightly better astrometric performance than RACS-Low1, using a common uncertainty facilitates a consistent comparison between the different reference catalogues. These examples demonstrate that even sub-arcsecond changes in localisation can have meaningful scientific implications, especially in cases where the host galaxy redshift is very high or when multiple potential hosts lie within the uncertainty region. While the corrected RACS-Low1 catalogue already provides a highly reliable reference, the adoption of RACS-Mid1 and RACS-High1 corrections for bursts detected in their respective bands ensures that systematic frequency-dependent offsets are minimised. This not only strengthens the robustness of host galaxy associations but also improves confidence in subsequent multiwavelength characterisation of FRB environments. Consequently, incorporating these higher-frequency catalogues into CELEBI represents a valuable advancement in the precision localisation of ASKAP FRBs.

\subsection{Future Plans} \label{sec:futureplans}

The next steps of this research will focus on addressing the persistent challenges of astrometric correction within the Galactic plane, where the high density of sources, prevalence of diffuse and extended emission, and the limitations of using WISE as a reference catalogue continue to degrade correction fidelity. The Galactic plane is especially problematic due to source blending and centroid biases caused by complex radio morphologies that differ substantially from their infrared counterparts. To overcome these issues, we propose to establish an independent astrometric reference specifically for low Galactic latitudes through phase-referenced observations with high-precision interferometers such as the Australia Telescope Compact Array (ATCA) and VLA (where available). By using these targeted observations to obtain accurate positions for compact, unresolved point sources in the plane, we can develop a dedicated correction framework that bypasses WISE entirely, thereby mitigating confusion and improving reliability in regions of high sky complexity.

In parallel, we plan to perform a unified cross-epoch analysis of the corrected RACS catalogues to identify persistent outliers or systematic discrepancies that remain after the present corrections. Such a comparison will help isolate sources with repeatable positional deviations and provide a basis for refining the global astrometric solution. This step is particularly important because RACS-Low1 currently serves as the reference layer for first-stage crossmatching in subsequent correction pipelines. Residual errors larger than $\sim1''$ at this stage can propagate and degrade the performance of later matching steps that operate with a $2''$ radius. Strengthening the internal consistency of RACS-Low1 is therefore essential to maintaining the precision of the broader correction framework.

A natural extension of this effort is to reprocess the per-beam RACS-Low3 source lists using the updated methodology introduced for the mid- and high-frequency epochs. The improved crossmatching hierarchy and enhanced modelling of sky-dependent systematics are expected to yield a low-frequency solution that surpasses the current RACS-Low1 reference in uniformity and robustness. Applying a single correction philosophy across all RACS epochs would reduce cross-epoch inconsistencies and provide a cleaner astrometric foundation for future catalogues, particularly given the central role of the low-frequency data in multi-stage localisation workflows. 

Together, these developments will move the RACS catalogues toward a spatially uniform, cross-band astrometric standard, ensuring that ASKAP continues to deliver reliable sub-arcsecond positions across the southern sky.

\section{CONCLUSION} \label{sec:conclusion}

This work extends our astrometric correction framework to the RACS-Mid1 and RACS-High1 epochs, completing the next major step toward a uniform high-precision reference across ASKAP’s operational frequency range. Building on lessons learned from RACS-Low3, we adopted a methodology that operates exclusively on unmosaicked per-beam source lists, thereby avoiding the positional biases introduced by mosaicking at beam boundaries that affected earlier public catalogues. Combined with a refined crossmatching hierarchy and improved modelling of sky-dependent systematics, this approach delivers stable and internally consistent corrections across the survey footprint. For both RACS-Mid1 and RACS-high1, the resulting astrometry achieves median residual offsets consistent with zero. These results demonstrate that the correction framework generalises effectively to higher-frequency observations while preserving the precision established in the low-frequency epoch, especially for declinations below $+30^\circ$.


Since the residual error distributions are shown to be exhibiting a longer tail than expected for a Gaussian distribution, we recommend using the following confidence intervals quantified from the WISE, FIRST and RFC comparisons. For both the RACS-Mid1 and RACS-High1 scans below Dec. $+30^\circ$, the standard deviation of the error distribution is at or slightly below $\pm0.25''$ for both RA and Dec. off the Galactic plane, and is at or slightly below $\pm0.35$ for both RA and Dec. on the plane. We accordingly recommend using these values by default, which will generally provide a conservative estimate over most of the sky, but suggest checking the sky distribution of WISE errors (Figures~\ref{fig:racsmid1aitoff} and \ref{fig:racshigh1aitoff} of this paper) to ascertain the possibility of a larger error for individual directions.



The corrected RACS-Mid1 and RACS-High1 catalogues represent the first sub-arcsecond astrometric reference across the southern sky above 1~GHz, improving positional accuracy by more than a factor of five compared to earlier southern surveys such as SUMSS which could only provide astrometric accuracy at the $\sim1''$--$2''$ level. Together with the corrected low-frequency catalogue, RACS now forms a coherent, cross-band astrometric standard that supports a wide range of ASKAP science. In particular, the improved background reference positions directly enhance the localisation of FRBs detected across ASKAP’s full bandwidth, strengthening host galaxy associations and enabling more reliable environmental and cosmological studies. The same precision benefits the localisation of other radio transients and variable sources, including flare stars and ultra-long-period transients, where accurate cross-identification with crowded optical catalogues is essential.

Collectively, these corrections establish RACS as the optimal astrometric reference catalogue in the southern hemisphere, complementing northern surveys such as VLASS and ensuring consistent positional standards across both hemispheres. This unified foundation is critical not only for current ASKAP science but also for future large-scale surveys, providing a stable astrometric backbone for transient discovery, galaxy evolution studies, and SKA-era radio astronomy.

\begin{acknowledgement}

AJ, ATD, and YW acknowledge the support of the Australian Research Council (ARC) grant DP220102305. MG is supported through UK STFC Grant ST/Y001117/1. MG acknowledges support from the Inter-University Institute for Data Intensive Astronomy (IDIA). IDIA is a partnership of the University of Cape Town, the University of Pretoria and the University of the Western Cape. For the purpose of open access, the author has applied a Creative Commons Attribution (CC BY) licence to any Author Accepted Manuscript version arising from this submission. We acknowledge the use of OpenAI's ChatGPT-5 (https://chatgpt.com/) and Google's NotebookLM (https://notebooklm.google.com/) for providing suggestions on refining sentence structure in this manuscript. This scientific work uses data obtained from Inyarrimanha Ilgari Bundara / the Murchison Radio-astronomy Observatory. We acknowledge the Wajarri Yamaji People as the Traditional Owners and native title holders of the Observatory site. CSIRO’s ASKAP radio telescope is part of the Australia Telescope National Facility (https://ror.org/05qajvd42). Operation of ASKAP is funded by the Australian Government with support from the National Collaborative Research Infrastructure Strategy. ASKAP uses the resources of the Pawsey Supercomputing Research Centre. Establishment of ASKAP, Inyarrimanha Ilgari Bundara, the CSIRO Murchison Radio-astronomy Observatory and the Pawsey Supercomputing Research Centre are initiatives of the Australian Government, with support from the Government of Western Australia and the Science and Industry Endowment Fund. This paper includes archived data obtained through the CSIRO ASKAP Science Data Archive, CASDA (https://data.csiro.au). A part of this work was performed on the OzSTAR national facility at Swinburne University of Technology. The OzSTAR programme receives funding in part from the Astronomy National Collaborative Research Infrastructure Strategy (NCRIS) allocation provided by the Australian Government. This research has made use of the VizieR catalogue access tool, CDS, Strasbourg, France \citealp{10.26093/cds/vizier}. The original description of the VizieR service was published in \citealp{vizier2000}. The Canadian Initiative for Radio Astronomy Data Analysis (CIRADA) is funded by a grant from the Canada Foundation for Innovation 2017 Innovation Fund (Project 35999) and by the Provinces of Ontario, British Columbia, Alberta, Manitoba and Quebec, in collaboration with the National Research Council of Canada, the US National Radio Astronomy Observatory and Australia’s Commonwealth Scientific and Industrial Research Organisation.


\end{acknowledgement}

\paragraph{Data Availability Statement}

The corrected RACS-Mid1 and RACS-High1 datasets, based on the methodology described in this work, are provided as beam-wise, unmosaicked source lists in VOTable (VOT) format through the PASA Datastore: \url{https://doi.org/10.57891/z4hj-8r90}. These datasets are intended to support the broader astronomical community in conducting high-precision localisation and cross-matching studies. Further details, including file structure and usage notes, are provided within the dataset descriptions on the PASA Datastore page.


\bibliography{ref}

@ARTICLE{2020PASA...37...48M,
       author = {{McConnell}, D. and {Hale}, C.~L. and {Lenc}, E. and {Banfield}, J.~K. and {Heald}, George and {Hotan}, A.~W. and {Leung}, James K. and {Moss}, Vanessa A. and {Murphy}, Tara and {O'Brien}, Andrew and {Pritchard}, Joshua and {Raja}, Wasim and {Sadler}, Elaine M. and {Stewart}, Adam and {Thomson}, Alec J.~M. and {Whiting}, M. and {Allison}, James R. and {Amy}, S.~W. and {Anderson}, C. and {Ball}, Lewis and {Bannister}, Keith W. and {Bell}, Martin and {Bock}, Douglas C. -J. and {Bolton}, Russ and {Bunton}, J.~D. and {Chippendale}, A.~P. and {Collier}, J.~D. and {Cooray}, F.~R. and {Cornwell}, T.~J. and {Diamond}, P.~J. and {Edwards}, P.~G. and {Gupta}, N. and {Hayman}, Douglas B. and {Heywood}, Ian and {Jackson}, C.~A. and {Koribalski}, B{\"a}rbel S. and {Lee-Waddell}, Karen and {McClure-Griffiths}, N.~M. and {Ng}, Alan and {Norris}, Ray P. and {Phillips}, Chris and {Reynolds}, John E. and {Roxby}, Daniel N. and {Schinckel}, Antony E.~T. and {Shields}, Matt and {Tremblay}, Chenoa and {Tzioumis}, A. and {Voronkov}, M.~A. and {Westmeier}, Tobias},
        title = "{The Rapid ASKAP Continuum Survey I: Design and first results}",
      journal = {\pasa},
         year = 2020,
        month = nov,
       volume = {37},
          eid = {e048},
        pages = {e048},
          doi = {10.1017/pasa.2020.41},
archivePrefix = {arXiv},
       eprint = {2012.00747},
 primaryClass = {astro-ph.IM},
       adsurl = {https://ui.adsabs.harvard.edu/abs/2020PASA...37...48M}
}

@ARTICLE{2023PASA...40...34D,
       author = {{Duchesne}, S.~W. and {Thomson}, A.~J.~M. and {Pritchard}, J. and {Lenc}, E. and {Moss}, V.~A. and {McConnell}, D. and {Wieringa}, M.~H. and {Whiting}, M.~T. and {Wang}, Z. and {Wang}, Y. and {Rose}, K. and {Raja}, W. and {Murphy}, Tara and {Leung}, J.~K. and {Huynh}, M.~T. and {Hotan}, A.~W. and {Hodgson}, T. and {Heald}, G.~H.},
        title = "{The Rapid ASKAP Continuum Survey IV: continuum imaging at 1367.5 MHz and the first data release of RACS-mid}",
      journal = {\pasa},
         year = 2023,
        month = aug,
       volume = {40},
          eid = {e034},
        pages = {e034},
          doi = {10.1017/pasa.2023.31},
archivePrefix = {arXiv},
       eprint = {2306.07194},
 primaryClass = {astro-ph.IM},
       adsurl = {https://ui.adsabs.harvard.edu/abs/2023PASA...40...34D}
}

@ARTICLE{2021PASA...38...58H,
       author = {{Hale}, Catherine L. and {McConnell}, D. and {Thomson}, A.~J.~M. and {Lenc}, E. and {Heald}, G.~H. and {Hotan}, A.~W. and {Leung}, J.~K. and {Moss}, V.~A. and {Murphy}, T. and {Pritchard}, J. and {Sadler}, E.~M. and {Stewart}, A.~J. and {Whiting}, M.~T.},
        title = "{The Rapid ASKAP Continuum Survey Paper II: First Stokes I Source Catalogue Data Release}",
      journal = {\pasa},
         year = 2021,
        month = nov,
       volume = {38},
          eid = {e058},
        pages = {e058},
          doi = {10.1017/pasa.2021.47},
archivePrefix = {arXiv},
       eprint = {2109.00956},
 primaryClass = {astro-ph.GA},
       adsurl = {https://ui.adsabs.harvard.edu/abs/2021PASA...38...58H}
}

@ARTICLE{2010AJ....140.1868W,
       author = {{Wright}, Edward L. and {Eisenhardt}, Peter R.~M. and {Mainzer}, Amy K. and {Ressler}, Michael E. and {Cutri}, Roc M. and {Jarrett}, Thomas and {Kirkpatrick}, J. Davy and {Padgett}, Deborah and {McMillan}, Robert S. and {Skrutskie}, Michael and {Stanford}, S.~A. and {Cohen}, Martin and {Walker}, Russell G. and {Mather}, John C. and {Leisawitz}, David and {Gautier}, Thomas N., III and {McLean}, Ian and {Benford}, Dominic and {Lonsdale}, Carol J. and {Blain}, Andrew and {Mendez}, Bryan and {Irace}, William R. and {Duval}, Valerie and {Liu}, Fengchuan and {Royer}, Don and {Heinrichsen}, Ingolf and {Howard}, Joan and {Shannon}, Mark and {Kendall}, Martha and {Walsh}, Amy L. and {Larsen}, Mark and {Cardon}, Joel G. and {Schick}, Scott and {Schwalm}, Mark and {Abid}, Mohamed and {Fabinsky}, Beth and {Naes}, Larry and {Tsai}, Chao-Wei},
        title = "{The Wide-field Infrared Survey Explorer (WISE): Mission Description and Initial On-orbit Performance}",
      journal = {\aj},
         year = 2010,
        month = dec,
       volume = {140},
       number = {6},
        pages = {1868-1881},
          doi = {10.1088/0004-6256/140/6/1868},
archivePrefix = {arXiv},
       eprint = {1008.0031},
 primaryClass = {astro-ph.IM},
       adsurl = {https://ui.adsabs.harvard.edu/abs/2010AJ....140.1868W}
}

@INPROCEEDINGS{1994ASPC...61..165B,
       author = {{Becker}, Robert H. and {White}, Richard L. and {Helfand}, David J.},
        title = "{The VLA's FIRST Survey}",
    booktitle = {Astronomical Data Analysis Software and Systems III},
         year = 1994,
       editor = {{Crabtree}, D.~R. and {Hanisch}, R.~J. and {Barnes}, J.},
       series = {Astronomical Society of the Pacific Conference Series},
       volume = {61},
        month = jan,
        pages = {165},
       adsurl = {https://ui.adsabs.harvard.edu/abs/1994ASPC...61..165B}
}

@ARTICLE{2020PASP..132c5001L,
       author = {{Lacy}, M. and {Baum}, S.~A. and {Chandler}, C.~J. and {Chatterjee}, S. and {Clarke}, T.~E. and {Deustua}, S. and {English}, J. and {Farnes}, J. and {Gaensler}, B.~M. and {Gugliucci}, N. and {Hallinan}, G. and {Kent}, B.~R. and {Kimball}, A. and {Law}, C.~J. and {Lazio}, T.~J.~W. and {Marvil}, J. and {Mao}, S.~A. and {Medlin}, D. and {Mooley}, K. and {Murphy}, E.~J. and {Myers}, S. and {Osten}, R. and {Richards}, G.~T. and {Rosolowsky}, E. and {Rudnick}, L. and {Schinzel}, F. and {Sivakoff}, G.~R. and {Sjouwerman}, L.~O. and {Taylor}, R. and {White}, R.~L. and {Wrobel}, J. and {Andernach}, H. and {Beasley}, A.~J. and {Berger}, E. and {Bhatnager}, S. and {Birkinshaw}, M. and {Bower}, G.~C. and {Brandt}, W.~N. and {Brown}, S. and {Burke-Spolaor}, S. and {Butler}, B.~J. and {Comerford}, J. and {Demorest}, P.~B. and {Fu}, H. and {Giacintucci}, S. and {Golap}, K. and {G{\"u}th}, T. and {Hales}, C.~A. and {Hiriart}, R. and {Hodge}, J. and {Horesh}, A. and {Ivezi{\'c}}, {\v{Z}}. and {Jarvis}, M.~J. and {Kamble}, A. and {Kassim}, N. and {Liu}, X. and {Loinard}, L. and {Lyons}, D.~K. and {Masters}, J. and {Mezcua}, M. and {Moellenbrock}, G.~A. and {Mroczkowski}, T. and {Nyland}, K. and {O'Dea}, C.~P. and {O'Sullivan}, S.~P. and {Peters}, W.~M. and {Radford}, K. and {Rao}, U. and {Robnett}, J. and {Salcido}, J. and {Shen}, Y. and {Sobotka}, A. and {Witz}, S. and {Vaccari}, M. and {van Weeren}, R.~J. and {Vargas}, A. and {Williams}, P.~K.~G. and {Yoon}, I.},
        title = "{The Karl G. Jansky Very Large Array Sky Survey (VLASS). Science Case and Survey Design}",
      journal = {\pasp},
         year = 2020,
        month = mar,
       volume = {132},
       number = {1009},
          eid = {035001},
        pages = {035001},
          doi = {10.1088/1538-3873/ab63eb},
archivePrefix = {arXiv},
       eprint = {1907.01981},
 primaryClass = {astro-ph.IM},
       adsurl = {https://ui.adsabs.harvard.edu/abs/2020PASP..132c5001L}
}

@ARTICLE{2023A&C....4400724S,
       author = {{Scott}, D.~R. and {Cho}, H. and {Day}, C.~K. and {Deller}, A.~T. and {Glowacki}, M. and {Gourdji}, K. and {Bannister}, K.~W. and {Bera}, A. and {Bhandari}, S. and {James}, C.~W. and {Shannon}, R.~M.},
        title = "{CELEBI: The CRAFT Effortless Localisation and Enhanced Burst Inspection pipeline}",
      journal = {Astronomy and Computing},
         year = 2023,
        month = jul,
       volume = {44},
          eid = {100724},
        pages = {100724},
          doi = {10.1016/j.ascom.2023.100724},
archivePrefix = {arXiv},
       eprint = {2301.13484},
 primaryClass = {astro-ph.IM},
       adsurl = {https://ui.adsabs.harvard.edu/abs/2023A&C....4400724S}
}

@ARTICLE{1998AJ....115.1693C,
       author = {{Condon}, J.~J. and {Cotton}, W.~D. and {Greisen}, E.~W. and {Yin}, Q.~F. and {Perley}, R.~A. and {Taylor}, G.~B. and {Broderick}, J.~J.},
        title = "{The NRAO VLA Sky Survey}",
      journal = {\aj},
         year = 1998,
        month = may,
       volume = {115},
       number = {5},
        pages = {1693-1716},
          doi = {10.1086/300337},
       adsurl = {https://ui.adsabs.harvard.edu/abs/1998AJ....115.1693C}
}

@ARTICLE{2021ApJS..255...30G,
       author = {{Gordon}, Yjan A. and {Boyce}, Michelle M. and {O'Dea}, Christopher P. and {Rudnick}, Lawrence and {Andernach}, Heinz and {Vantyghem}, Adrian N. and {Baum}, Stefi A. and {Bui}, Jean-Paul and {Dionyssiou}, Mathew and {Safi-Harb}, Samar and {Sander}, Isabel},
        title = "{A Quick Look at the 3 GHz Radio Sky. I. Source Statistics from the Very Large Array Sky Survey}",
      journal = {\apjs},
         year = 2021,
        month = aug,
       volume = {255},
       number = {2},
          eid = {30},
        pages = {30},
          doi = {10.3847/1538-4365/ac05c0},
archivePrefix = {arXiv},
       eprint = {2102.11753},
 primaryClass = {astro-ph.GA},
       adsurl = {https://ui.adsabs.harvard.edu/abs/2021ApJS..255...30G}
}

@ARTICLE{2010PASA...27..272M,
       author = {{Macquart}, Jean-Pierre and {Bailes}, M. and {Bhat}, N.~D.~R. and {Bower}, G.~C. and {Bunton}, J.~D. and {Chatterjee}, S. and {Colegate}, T. and {Cordes}, J.~M. and {D'Addario}, L. and {Deller}, A. and {Dodson}, R. and {Fender}, R. and {Haines}, K. and {Halll}, P. and {Harris}, C. and {Hotan}, A. and {Johnston}, S. and {Jones}, D.~L. and {Keith}, M. and {Koay}, J.~Y. and {Lazio}, T.~J.~W. and {Majid}, W. and {Murphy}, T. and {Navarro}, R. and {Phillips}, C. and {Quinn}, P. and {Preston}, R.~A. and {Stansby}, B. and {Stairs}, I. and {Stappers}, B. and {Staveley-Smith}, L. and {Tingay}, S. and {Thompson}, D. and {van Straten}, W. and {Wagstaff}, K. and {Warren}, M. and {Wayth}, R. and {Wen}, L. and {CRAFT Collaboration}},
        title = "{The Commensal Real-Time ASKAP Fast-Transients (CRAFT) Survey}",
      journal = {\pasa},
         year = 2010,
        month = jun,
       volume = {27},
       number = {3},
        pages = {272-282},
          doi = {10.1071/AS09082},
archivePrefix = {arXiv},
       eprint = {1001.2958},
 primaryClass = {astro-ph.HE},
       adsurl = {https://ui.adsabs.harvard.edu/abs/2010PASA...27..272M}
}

@ARTICLE{2021PASA...38....9H,
       author = {{Hotan}, A.~W. and {Bunton}, J.~D. and {Chippendale}, A.~P. and {Whiting}, M. and {Tuthill}, J. and {Moss}, V.~A. and {McConnell}, D. and {Amy}, S.~W. and {Huynh}, M.~T. and {Allison}, J.~R. and {Anderson}, C.~S. and {Bannister}, K.~W. and {Bastholm}, E. and {Beresford}, R. and {Bock}, D.~C. -J. and {Bolton}, R. and {Chapman}, J.~M. and {Chow}, K. and {Collier}, J.~D. and {Cooray}, F.~R. and {Cornwell}, T.~J. and {Diamond}, P.~J. and {Edwards}, P.~G. and {Feain}, I.~J. and {Franzen}, T.~M.~O. and {George}, D. and {Gupta}, N. and {Hampson}, G.~A. and {Harvey-Smith}, L. and {Hayman}, D.~B. and {Heywood}, I. and {Jacka}, C. and {Jackson}, C.~A. and {Jackson}, S. and {Jeganathan}, K. and {Johnston}, S. and {Kesteven}, M. and {Kleiner}, D. and {Koribalski}, B.~S. and {Lee-Waddell}, K. and {Lenc}, E. and {Lensson}, E.~S. and {Mackay}, S. and {Mahony}, E.~K. and {McClure-Griffiths}, N.~M. and {McConigley}, R. and {Mirtschin}, P. and {Ng}, A.~K. and {Norris}, R.~P. and {Pearce}, S.~E. and {Phillips}, C. and {Pilawa}, M.~A. and {Raja}, W. and {Reynolds}, J.~E. and {Roberts}, P. and {Roxby}, D.~N. and {Sadler}, E.~M. and {Shields}, M. and {Schinckel}, A.~E.~T. and {Serra}, P. and {Shaw}, R.~D. and {Sweetnam}, T. and {Troup}, E.~R. and {Tzioumis}, A. and {Voronkov}, M.~A. and {Westmeier}, T.},
        title = "{Australian square kilometre array pathfinder: I. system description}",
      journal = {\pasa},
         year = 2021,
        month = mar,
       volume = {38},
          eid = {e009},
        pages = {e009},
          doi = {10.1017/pasa.2021.1},
archivePrefix = {arXiv},
       eprint = {2102.01870},
 primaryClass = {astro-ph.IM},
       adsurl = {https://ui.adsabs.harvard.edu/abs/2021PASA...38....9H}
}

@ARTICLE{2024PASA...41....3D,
       author = {{Duchesne}, S.~W. and {Grundy}, J.~A. and {Heald}, George H. and {Lenc}, Emil and {Leung}, James K. and {McConnell}, David and {Murphy}, Tara and {Pritchard}, Joshua and {Rose}, Kovi and {Thomson}, Alec J.~M. and {Wang}, Yuanming and {Wang}, Ziteng and {Whiting}, Matthew T.},
        title = "{The Rapid ASKAP Continuum Survey V: Cataloguing the sky at 1 367.5 MHz and the second data release of RACS-mid}",
      journal = {\pasa},
         year = 2024,
        month = jan,
       volume = {41},
          eid = {e003},
        pages = {e003},
          doi = {10.1017/pasa.2023.60},
archivePrefix = {arXiv},
       eprint = {2311.12369},
 primaryClass = {astro-ph.GA},
       adsurl = {https://ui.adsabs.harvard.edu/abs/2024PASA...41....3D}
}

@ARTICLE{2020SciPy-NMeth,
  author  = {Virtanen, Pauli and Gommers, Ralf and Oliphant, Travis E. and
            Haberland, Matt and Reddy, Tyler and Cournapeau, David and
            Burovski, Evgeni and Peterson, Pearu and Weckesser, Warren and
            Bright, Jonathan and {van der Walt}, St{\'e}fan J. and
            Brett, Matthew and Wilson, Joshua and Millman, K. Jarrod and
            Mayorov, Nikolay and Nelson, Andrew R. J. and Jones, Eric and
            Kern, Robert and Larson, Eric and Carey, C J and
            Polat, {\.I}lhan and Feng, Yu and Moore, Eric W. and
            {VanderPlas}, Jake and Laxalde, Denis and Perktold, Josef and
            Cimrman, Robert and Henriksen, Ian and Quintero, E. A. and
            Harris, Charles R. and Archibald, Anne M. and
            Ribeiro, Ant{\^o}nio H. and Pedregosa, Fabian and
            {van Mulbregt}, Paul and {SciPy 1.0 Contributors}},
  title   = {{{SciPy} 1.0: Fundamental Algorithms for Scientific
            Computing in Python}},
  journal = {Nature Methods},
  year    = {2020},
  volume  = {17},
  pages   = {261--272},
  adsurl  = {https://rdcu.be/b08Wh},
  doi     = {10.1038/s41592-019-0686-2},
}

@article{astropy:2013,
Adsurl = {http://adsabs.harvard.edu/abs/2013A%26A...558A..33A},
Archiveprefix = {arXiv},
Author = {{Astropy Collaboration} and {Robitaille}, T.~P. and {Tollerud}, E.~J. and {Greenfield}, P. and {Droettboom}, M. and {Bray}, E. and {Aldcroft}, T. and {Davis}, M. and {Ginsburg}, A. and {Price-Whelan}, A.~M. and {Kerzendorf}, W.~E. and {Conley}, A. and {Crighton}, N. and {Barbary}, K. and {Muna}, D. and {Ferguson}, H. and {Grollier}, F. and {Parikh}, M.~M. and {Nair}, P.~H. and {Unther}, H.~M. and {Deil}, C. and {Woillez}, J. and {Conseil}, S. and {Kramer}, R. and {Turner}, J.~E.~H. and {Singer}, L. and {Fox}, R. and {Weaver}, B.~A. and {Zabalza}, V. and {Edwards}, Z.~I. and {Azalee Bostroem}, K. and {Burke}, D.~J. and {Casey}, A.~R. and {Crawford}, S.~M. and {Dencheva}, N. and {Ely}, J. and {Jenness}, T. and {Labrie}, K. and {Lim}, P.~L. and {Pierfederici}, F. and {Pontzen}, A. and {Ptak}, A. and {Refsdal}, B. and {Servillat}, M. and {Streicher}, O.},
Doi = {10.1051/0004-6361/201322068},
Eid = {A33},
Eprint = {1307.6212},
Journal = {\aap},
Month = oct,
Pages = {A33},
Primaryclass = {astro-ph.IM},
Title = {{Astropy: A community Python package for astronomy}},
Volume = 558,
Year = 2013}

@ARTICLE{astropy:2018,
       author = {{Astropy Collaboration} and {Price-Whelan}, A.~M. and
         {Sip{\H{o}}cz}, B.~M. and {G{\"u}nther}, H.~M. and {Lim}, P.~L. and
         {Crawford}, S.~M. and {Conseil}, S. and {Shupe}, D.~L. and
         {Craig}, M.~W. and {Dencheva}, N. and {Ginsburg}, A. and {Vand
        erPlas}, J.~T. and {Bradley}, L.~D. and {P{\'e}rez-Su{\'a}rez}, D. and
         {de Val-Borro}, M. and {Aldcroft}, T.~L. and {Cruz}, K.~L. and
         {Robitaille}, T.~P. and {Tollerud}, E.~J. and {Ardelean}, C. and
         {Babej}, T. and {Bach}, Y.~P. and {Bachetti}, M. and {Bakanov}, A.~V. and
         {Bamford}, S.~P. and {Barentsen}, G. and {Barmby}, P. and
         {Baumbach}, A. and {Berry}, K.~L. and {Biscani}, F. and {Boquien}, M. and
         {Bostroem}, K.~A. and {Bouma}, L.~G. and {Brammer}, G.~B. and
         {Bray}, E.~M. and {Breytenbach}, H. and {Buddelmeijer}, H. and
         {Burke}, D.~J. and {Calderone}, G. and {Cano Rodr{\'\i}guez}, J.~L. and
         {Cara}, M. and {Cardoso}, J.~V.~M. and {Cheedella}, S. and {Copin}, Y. and
         {Corrales}, L. and {Crichton}, D. and {D'Avella}, D. and {Deil}, C. and
         {Depagne}, {\'E}. and {Dietrich}, J.~P. and {Donath}, A. and
         {Droettboom}, M. and {Earl}, N. and {Erben}, T. and {Fabbro}, S. and
         {Ferreira}, L.~A. and {Finethy}, T. and {Fox}, R.~T. and
         {Garrison}, L.~H. and {Gibbons}, S.~L.~J. and {Goldstein}, D.~A. and
         {Gommers}, R. and {Greco}, J.~P. and {Greenfield}, P. and
         {Groener}, A.~M. and {Grollier}, F. and {Hagen}, A. and {Hirst}, P. and
         {Homeier}, D. and {Horton}, A.~J. and {Hosseinzadeh}, G. and {Hu}, L. and
         {Hunkeler}, J.~S. and {Ivezi{\'c}}, {\v{Z}}. and {Jain}, A. and
         {Jenness}, T. and {Kanarek}, G. and {Kendrew}, S. and {Kern}, N.~S. and
         {Kerzendorf}, W.~E. and {Khvalko}, A. and {King}, J. and {Kirkby}, D. and
         {Kulkarni}, A.~M. and {Kumar}, A. and {Lee}, A. and {Lenz}, D. and
         {Littlefair}, S.~P. and {Ma}, Z. and {Macleod}, D.~M. and
         {Mastropietro}, M. and {McCully}, C. and {Montagnac}, S. and
         {Morris}, B.~M. and {Mueller}, M. and {Mumford}, S.~J. and {Muna}, D. and
         {Murphy}, N.~A. and {Nelson}, S. and {Nguyen}, G.~H. and
         {Ninan}, J.~P. and {N{\"o}the}, M. and {Ogaz}, S. and {Oh}, S. and
         {Parejko}, J.~K. and {Parley}, N. and {Pascual}, S. and {Patil}, R. and
         {Patil}, A.~A. and {Plunkett}, A.~L. and {Prochaska}, J.~X. and
         {Rastogi}, T. and {Reddy Janga}, V. and {Sabater}, J. and
         {Sakurikar}, P. and {Seifert}, M. and {Sherbert}, L.~E. and
         {Sherwood-Taylor}, H. and {Shih}, A.~Y. and {Sick}, J. and
         {Silbiger}, M.~T. and {Singanamalla}, S. and {Singer}, L.~P. and
         {Sladen}, P.~H. and {Sooley}, K.~A. and {Sornarajah}, S. and
         {Streicher}, O. and {Teuben}, P. and {Thomas}, S.~W. and
         {Tremblay}, G.~R. and {Turner}, J.~E.~H. and {Terr{\'o}n}, V. and
         {van Kerkwijk}, M.~H. and {de la Vega}, A. and {Watkins}, L.~L. and
         {Weaver}, B.~A. and {Whitmore}, J.~B. and {Woillez}, J. and
         {Zabalza}, V. and {Astropy Contributors}},
        title = "{The Astropy Project: Building an Open-science Project and Status of the v2.0 Core Package}",
      journal = {\aj},
         year = 2018,
        month = sep,
       volume = {156},
       number = {3},
          eid = {123},
        pages = {123},
          doi = {10.3847/1538-3881/aabc4f},
archivePrefix = {arXiv},
       eprint = {1801.02634},
 primaryClass = {astro-ph.IM},
       adsurl = {https://ui.adsabs.harvard.edu/abs/2018AJ....156..123A}
}

@ARTICLE{astropy:2022,
       author = {{Astropy Collaboration} and {Price-Whelan}, Adrian M. and {Lim}, Pey Lian and {Earl}, Nicholas and {Starkman}, Nathaniel and {Bradley}, Larry and {Shupe}, David L. and {Patil}, Aarya A. and {Corrales}, Lia and {Brasseur}, C.~E. and {N{"o}the}, Maximilian and {Donath}, Axel and {Tollerud}, Erik and {Morris}, Brett M. and {Ginsburg}, Adam and {Vaher}, Eero and {Weaver}, Benjamin A. and {Tocknell}, James and {Jamieson}, William and {van Kerkwijk}, Marten H. and {Robitaille}, Thomas P. and {Merry}, Bruce and {Bachetti}, Matteo and {G{"u}nther}, H. Moritz and {Aldcroft}, Thomas L. and {Alvarado-Montes}, Jaime A. and {Archibald}, Anne M. and {B{'o}di}, Attila and {Bapat}, Shreyas and {Barentsen}, Geert and {Baz{'a}n}, Juanjo and {Biswas}, Manish and {Boquien}, M{'e}d{'e}ric and {Burke}, D.~J. and {Cara}, Daria and {Cara}, Mihai and {Conroy}, Kyle E. and {Conseil}, Simon and {Craig}, Matthew W. and {Cross}, Robert M. and {Cruz}, Kelle L. and {D'Eugenio}, Francesco and {Dencheva}, Nadia and {Devillepoix}, Hadrien A.~R. and {Dietrich}, J{"o}rg P. and {Eigenbrot}, Arthur Davis and {Erben}, Thomas and {Ferreira}, Leonardo and {Foreman-Mackey}, Daniel and {Fox}, Ryan and {Freij}, Nabil and {Garg}, Suyog and {Geda}, Robel and {Glattly}, Lauren and {Gondhalekar}, Yash and {Gordon}, Karl D. and {Grant}, David and {Greenfield}, Perry and {Groener}, Austen M. and {Guest}, Steve and {Gurovich}, Sebastian and {Handberg}, Rasmus and {Hart}, Akeem and {Hatfield-Dodds}, Zac and {Homeier}, Derek and {Hosseinzadeh}, Griffin and {Jenness}, Tim and {Jones}, Craig K. and {Joseph}, Prajwel and {Kalmbach}, J. Bryce and {Karamehmetoglu}, Emir and {Ka{l}uszy{'n}ski}, Miko{l}aj and {Kelley}, Michael S.~P. and {Kern}, Nicholas and {Kerzendorf}, Wolfgang E. and {Koch}, Eric W. and {Kulumani}, Shankar and {Lee}, Antony and {Ly}, Chun and {Ma}, Zhiyuan and {MacBride}, Conor and {Maljaars}, Jakob M. and {Muna}, Demitri and {Murphy}, N.~A. and {Norman}, Henrik and {O'Steen}, Richard and {Oman}, Kyle A. and {Pacifici}, Camilla and {Pascual}, Sergio and {Pascual-Granado}, J. and {Patil}, Rohit R. and {Perren}, Gabriel I. and {Pickering}, Timothy E. and {Rastogi}, Tanuj and {Roulston}, Benjamin R. and {Ryan}, Daniel F. and {Rykoff}, Eli S. and {Sabater}, Jose and {Sakurikar}, Parikshit and {Salgado}, Jes{'u}s and {Sanghi}, Aniket and {Saunders}, Nicholas and {Savchenko}, Volodymyr and {Schwardt}, Ludwig and {Seifert-Eckert}, Michael and {Shih}, Albert Y. and {Jain}, Anany Shrey and {Shukla}, Gyanendra and {Sick}, Jonathan and {Simpson}, Chris and {Singanamalla}, Sudheesh and {Singer}, Leo P. and {Singhal}, Jaladh and {Sinha}, Manodeep and {Sip{H{o}}cz}, Brigitta M. and {Spitler}, Lee R. and {Stansby}, David and {Streicher}, Ole and {{{S}}umak}, Jani and {Swinbank}, John D. and {Taranu}, Dan S. and {Tewary}, Nikita and {Tremblay}, Grant R. and {Val-Borro}, Miguel de and {Van Kooten}, Samuel J. and {Vasovi{'c}}, Zlatan and {Verma}, Shresth and {de Miranda Cardoso}, Jos{'e} Vin{'i}cius and {Williams}, Peter K.~G. and {Wilson}, Tom J. and {Winkel}, Benjamin and {Wood-Vasey}, W.~M. and {Xue}, Rui and {Yoachim}, Peter and {Zhang}, Chen and {Zonca}, Andrea and {Astropy Project Contributors}},
        title = "{The Astropy Project: Sustaining and Growing a Community-oriented Open-source Project and the Latest Major Release (v5.0) of the Core Package}",
      journal = {\apj},
         year = 2022,
        month = aug,
       volume = {935},
       number = {2},
          eid = {167},
        pages = {167},
          doi = {10.3847/1538-4357/ac7c74},
archivePrefix = {arXiv},
       eprint = {2206.14220},
 primaryClass = {astro-ph.IM},
       adsurl = {https://ui.adsabs.harvard.edu/abs/2022ApJ...935..167A}
}

@misc{10.26093/cds/vizier,
  doi = {10.26093/CDS/VIZIER},
  url = {https://vizier.cds.unistra.fr},
  author = {Ochsenbein,  Francois},
  title = {The VizieR database of astronomical catalogues},
  publisher = {CDS,  Centre de DonnÃ©es astronomiques de Strasbourg},
  year = {1996},
  copyright = {Refer to CDS usage}
}

@ARTICLE{vizier2000,
       author = {{Ochsenbein}, F. and {Bauer}, P. and {Marcout}, J.},
        title = "{The VizieR database of astronomical catalogues}",
      journal = {\aaps},
         year = 2000,
        month = apr,
       volume = {143},
        pages = {23-32},
          doi = {10.1051/aas:2000169},
archivePrefix = {arXiv},
       eprint = {astro-ph/0002122},
 primaryClass = {astro-ph},
       adsurl = {https://ui.adsabs.harvard.edu/abs/2000A&AS..143...23O}
}

@ARTICLE{2019AJ....157...98G,
       author = {{Ginsburg}, Adam and {Sip{\H{o}}cz}, Brigitta M. and {Brasseur}, C.~E. and {Cowperthwaite}, Philip S. and {Craig}, Matthew W. and {Deil}, Christoph and {Guillochon}, James and {Guzman}, Giannina and {Liedtke}, Simon and {Lian Lim}, Pey and {Lockhart}, Kelly E. and {Mommert}, Michael and {Morris}, Brett M. and {Norman}, Henrik and {Parikh}, Madhura and {Persson}, Magnus V. and {Robitaille}, Thomas P. and {Segovia}, Juan-Carlos and {Singer}, Leo P. and {Tollerud}, Erik J. and {de Val-Borro}, Miguel and {Valtchanov}, Ivan and {Woillez}, Julien and {Astroquery Collaboration} and {a subset of astropy Collaboration}},
        title = "{astroquery: An Astronomical Web-querying Package in Python}",
      journal = {\aj},
         year = 2019,
        month = mar,
       volume = {157},
       number = {3},
          eid = {98},
        pages = {98},
          doi = {10.3847/1538-3881/aafc33},
archivePrefix = {arXiv},
       eprint = {1901.04520},
 primaryClass = {astro-ph.IM},
       adsurl = {https://ui.adsabs.harvard.edu/abs/2019AJ....157...98G}
}

@Article{harris2020array,
 title         = {Array programming with {NumPy}},
 author        = {Charles R. Harris and K. Jarrod Millman and St{\'{e}}fan J.
                 van der Walt and Ralf Gommers and Pauli Virtanen and David
                 Cournapeau and Eric Wieser and Julian Taylor and Sebastian
                 Berg and Nathaniel J. Smith and Robert Kern and Matti Picus
                 and Stephan Hoyer and Marten H. van Kerkwijk and Matthew
                 Brett and Allan Haldane and Jaime Fern{\'{a}}ndez del
                 R{\'{i}}o and Mark Wiebe and Pearu Peterson and Pierre
                 G{\'{e}}rard-Marchant and Kevin Sheppard and Tyler Reddy and
                 Warren Weckesser and Hameer Abbasi and Christoph Gohlke and
                 Travis E. Oliphant},
 year          = {2020},
 month         = sep,
 journal       = {Nature},
 volume        = {585},
 number        = {7825},
 pages         = {357--362},
 doi           = {10.1038/s41586-020-2649-2},
 publisher     = {Springer Science and Business Media {LLC}},
 url           = {https://doi.org/10.1038/s41586-020-2649-2}
}

@article{jaini2025enhanced,
  title={Enhanced astrometry of the rapid ASKAP continuum survey for precise localisation of fast radio bursts},
  author={Jaini, Akhil and Deller, Adam T and Wang, Yuanming and Lenc, Emil and Glowacki, Marcin},
  journal={Publications of the Astronomical Society of Australia},
  volume={42},
  pages={e060},
  year={2025},
  publisher={Cambridge University Press}
}

@article{celebi2.0,
  title={A PINK update: Improvements to the CELEBI fast radio burst data reduction and analysis pipeline},
  author={Glowacki, M and Dial, T and Bera, A and Deller, AT and Gourdji, K and Jaini, A and Scott, D and Wang, Y and Desnos, K and Gordon, AC and others},
  journal={arXiv preprint arXiv:2605.06766},
  year={2026}
}

@ARTICLE{2021A&A...647L..11I,
       author = {{Ighina}, L. and {Belladitta}, S. and {Caccianiga}, A. and {Broderick}, J.~W. and {Drouart}, G. and {Moretti}, A. and {Seymour}, N.},
        title = "{Radio detection of VIK J2318{\ensuremath{-}}3113, the most distant radio-loud quasar (z = 6.44)}",
      journal = {\aap},
         year = 2021,
        month = mar,
       volume = {647},
          eid = {L11},
        pages = {L11},
          doi = {10.1051/0004-6361/202140362},
archivePrefix = {arXiv},
       eprint = {2101.11371},
 primaryClass = {astro-ph.GA},
       adsurl = {https://ui.adsabs.harvard.edu/abs/2021A&A...647L..11I}
}

@ARTICLE{2021Galax...9...99A,
       author = {{Andernach}, Heinz and {Jim{\'e}nez-Andrade}, Eric F. and {Willis}, Anthony G.},
        title = "{Discovery of 178 Giant Radio Galaxies in 1059 deg$^{2}$ of the Rapid ASKAP Continuum Survey at 888 MHz}",
      journal = {Galaxies},
         year = 2021,
        month = nov,
       volume = {9},
       number = {4},
          eid = {99},
        pages = {99},
          doi = {10.3390/galaxies9040099},
archivePrefix = {arXiv},
       eprint = {2111.08807},
 primaryClass = {astro-ph.GA},
       adsurl = {https://ui.adsabs.harvard.edu/abs/2021Galax...9...99A}
}

@ARTICLE{2022MNRAS.511.3525D,
       author = {{Duchesne}, S.~W. and {Johnston-Hollitt}, M. and {Riseley}, C.~J. and {Bartalucci}, I. and {Keel}, S.~R.},
        title = "{The merging galaxy cluster Abell 3266 at low radio frequencies}",
      journal = {\mnras},
         year = 2022,
        month = apr,
       volume = {511},
       number = {3},
        pages = {3525-3535},
          doi = {10.1093/mnras/stac335},
archivePrefix = {arXiv},
       eprint = {2201.09629},
 primaryClass = {astro-ph.CO},
       adsurl = {https://ui.adsabs.harvard.edu/abs/2022MNRAS.511.3525D}
}

@ARTICLE{2022NatAs...6..828C,
       author = {{Caleb}, Manisha and {Heywood}, Ian and {Rajwade}, Kaustubh and {Malenta}, Mateusz and {Stappers}, Benjamin Willem and {Barr}, Ewan and {Chen}, Weiwei and {Morello}, Vincent and {Sanidas}, Sotiris and {van den Eijnden}, Jakob and {Kramer}, Michael and {Buckley}, David and {Brink}, Jaco and {Motta}, Sara Elisa and {Woudt}, Patrick and {Weltevrede}, Patrick and {Jankowski}, Fabian and {Surnis}, Mayuresh and {Buchner}, Sarah and {Bezuidenhout}, Mechiel Christiaan and {Driessen}, Laura Nicole and {Fender}, Rob},
        title = "{Discovery of a radio-emitting neutron star with an ultra-long spin period of 76 s}",
      journal = {Nature Astronomy},
         year = 2022,
        month = may,
       volume = {6},
        pages = {828-836},
          doi = {10.1038/s41550-022-01688-x},
archivePrefix = {arXiv},
       eprint = {2206.01346},
 primaryClass = {astro-ph.HE},
       adsurl = {https://ui.adsabs.harvard.edu/abs/2022NatAs...6..828C}
}

@ARTICLE{2023ApJ...948...67B,
       author = {{Bhandari}, Shivani and {Gordon}, Alexa C. and {Scott}, Danica R. and {Marnoch}, Lachlan and {Sridhar}, Navin and {Kumar}, Pravir and {James}, Clancy W. and {Qiu}, Hao and {Bannister}, Keith W. and {T. Deller}, Adam and {Eftekhari}, Tarraneh and {Fong}, Wen-fai and {Glowacki}, Marcin and {Prochaska}, J. Xavier and {Ryder}, Stuart D. and {Shannon}, Ryan M. and {Simha}, Sunil},
        title = "{A Nonrepeating Fast Radio Burst in a Dwarf Host Galaxy}",
      journal = {\apj},
         year = 2023,
        month = may,
       volume = {948},
       number = {1},
          eid = {67},
        pages = {67},
          doi = {10.3847/1538-4357/acc178},
archivePrefix = {arXiv},
       eprint = {2211.16790},
 primaryClass = {astro-ph.HE},
       adsurl = {https://ui.adsabs.harvard.edu/abs/2023ApJ...948...67B}
}

@ARTICLE{2023ApJ...956...28A,
       author = {{Anumarlapudi}, Akash and {Ehlke}, Anna and {Jones}, Megan L. and {Kaplan}, David L. and {Dobie}, Dougal and {Lenc}, Emil and {Leung}, James K. and {Murphy}, Tara and {Pritchard}, Joshua and {Stewart}, Adam J. and {Sengar}, Rahul and {Anderson}, Craig and {Banfield}, Julie and {Heald}, George and {Hotan}, Aidan W. and {McConnell}, David and {Moss}, Vanessa A. and {Raja}, Wasim and {Whiting}, Matthew T.},
        title = "{Characterizing Pulsars Detected in the Rapid ASKAP Continuum Survey}",
      journal = {\apj},
         year = 2023,
        month = oct,
       volume = {956},
       number = {1},
          eid = {28},
        pages = {28},
          doi = {10.3847/1538-4357/aceb5d},
archivePrefix = {arXiv},
       eprint = {2308.00100},
 primaryClass = {astro-ph.HE},
       adsurl = {https://ui.adsabs.harvard.edu/abs/2023ApJ...956...28A}
}

@ARTICLE{2021PASA...38...50D,
       author = {{Day}, Cherie K. and {Deller}, Adam T. and {James}, Clancy W. and {Lenc}, Emil and {Bhandari}, Shivani and {Shannon}, R.~M. and {Bannister}, Keith W.},
        title = "{Astrometric accuracy of snapshot fast radio burst localisations with ASKAP}",
      journal = {\pasa},
         year = 2021,
        month = sep,
       volume = {38},
          eid = {e050},
        pages = {e050},
          doi = {10.1017/pasa.2021.40},
archivePrefix = {arXiv},
       eprint = {2107.07068},
 primaryClass = {astro-ph.IM},
       adsurl = {https://ui.adsabs.harvard.edu/abs/2021PASA...38...50D}
}

@ARTICLE{2023MNRAS.523.5661W,
       author = {{Wang}, Yuanming and {Murphy}, Tara and {Lenc}, Emil and {Mercorelli}, Louis and {Driessen}, Laura and {Pritchard}, Joshua and {Lao}, Baoqiang and {Kaplan}, David L. and {An}, Tao and {Bannister}, Keith W. and {Heald}, George and {Lu}, Shuoying and {Tuntsov}, Artem and {Walker}, Mark and {Zic}, Andrew},
        title = "{Radio variable and transient sources on minute time-scales in the ASKAP pilot surveys}",
      journal = {\mnras},
         year = 2023,
        month = aug,
       volume = {523},
       number = {4},
        pages = {5661-5680},
          doi = {10.1093/mnras/stad1727},
archivePrefix = {arXiv},
       eprint = {2306.04263},
 primaryClass = {astro-ph.HE},
       adsurl = {https://ui.adsabs.harvard.edu/abs/2023MNRAS.523.5661W}
}

@ARTICLE{2025A&A...698A.158I,
       author = {{Ighina}, L. and {Caccianiga}, A. and {Moretti}, A. and {Broderick}, J.~W. and {Leung}, J.~K. and {Rigamonti}, F. and {Seymour}, N. and {Afonso}, J. and {Connor}, T. and {Vignali}, C. and {Wang}, Z. and {An}, T. and {Arsioli}, B. and {Bisogni}, S. and {Dallacasa}, D. and {Della Ceca}, R. and {Liu}, Y. and {L{\'o}pez-S{\'a}nchez}, A. and {Matute}, I. and {Reynolds}, C. and {Rossi}, A. and {Spingola}, C. and {Severgnini}, P. and {Tavecchio}, F.},
        title = "{High-z radio quasars in RACS: I. Selection, identification, and multi-wavelength properties}",
      journal = {\aap},
         year = 2025,
        month = jun,
       volume = {698},
          eid = {A158},
        pages = {A158},
          doi = {10.1051/0004-6361/202453650},
archivePrefix = {arXiv},
       eprint = {2504.10573},
 primaryClass = {astro-ph.GA},
       adsurl = {https://ui.adsabs.harvard.edu/abs/2025A&A...698A.158I}
}

@ARTICLE{2024PASA...41....4B,
       author = {{Brown}, Michael J.~I. and {Clarke}, Teagan A. and {Hopkins}, Andrew M. and {Norris}, Ray P. and {Jarrett}, T.~H.},
        title = "{Radio continuum from the most massive early-type galaxies detected with ASKAP RACS}",
      journal = {\pasa},
         year = 2024,
        month = jan,
       volume = {41},
          eid = {e004},
        pages = {e004},
          doi = {10.1017/pasa.2023.62},
archivePrefix = {arXiv},
       eprint = {2311.15456},
 primaryClass = {astro-ph.GA},
       adsurl = {https://ui.adsabs.harvard.edu/abs/2024PASA...41....4B}
}

@article{duchesne2025rapid,
  title={The Rapid ASKAP Continuum Survey (RACS) VI: The RACS-high 1655.5 MHz images and catalogue},
  author={Duchesne, SW and Ross, K and Thomson, AJM and Lenc, E and Murphy, Tara and Galvin, TJ and Hotan, AW and Moss, VA and Whiting, Matthew T},
  journal={arXiv preprint arXiv:2501.04978},
  year={2025}
}

@ARTICLE{1999AJ....117.1578B,
       author = {{Bock}, D.~C.-J. and {Large}, M.~I. and {Sadler}, Elaine M.},
        title = "{SUMSS: A Wide-Field Radio Imaging Survey of the Southern Sky. I. Science Goals, Survey Design, and Instrumentation}",
      journal = {\aj},
         year = 1999,
        month = mar,
       volume = {117},
       number = {3},
        pages = {1578-1593},
          doi = {10.1086/300786},
archivePrefix = {arXiv},
       eprint = {astro-ph/9812083},
 primaryClass = {astro-ph},
       adsurl = {https://ui.adsabs.harvard.edu/abs/1999AJ....117.1578B}
}

@ARTICLE{2020A&A...644A.159C,
       author = {{Charlot}, P. and {Jacobs}, C.~S. and {Gordon}, D. and {Lambert}, S. and {de Witt}, A. and {B{\"o}hm}, J. and {Fey}, A.~L. and {Heinkelmann}, R. and {Skurikhina}, E. and {Titov}, O. and {Arias}, E.~F. and {Bolotin}, S. and {Bourda}, G. and {Ma}, C. and {Malkin}, Z. and {Nothnagel}, A. and {Mayer}, D. and {MacMillan}, D.~S. and {Nilsson}, T. and {Gaume}, R.},
        title = "{The third realization of the International Celestial Reference Frame by very long baseline interferometry}",
      journal = {\aap},
         year = 2020,
        month = dec,
       volume = {644},
          eid = {A159},
        pages = {A159},
          doi = {10.1051/0004-6361/202038368},
archivePrefix = {arXiv},
       eprint = {2010.13625},
 primaryClass = {astro-ph.GA},
       adsurl = {https://ui.adsabs.harvard.edu/abs/2020A&A...644A.159C}
}

@article{james2025nanosecond,
  title={A nanosecond-duration radio pulse originating from the defunct Relay 2 satellite},
  author={James, CW and Deller, AT and Dial, T and Glowacki, M and Tingay, SJ and Bannister, KW and Bera, A and Bhat, NDR and Ekers, RD and Gupta, V and others},
  journal={The Astrophysical Journal Letters},
  volume={987},
  number={1},
  pages={L16},
  year={2025},
  publisher={IOP Publishing}
}

@article{petrov2025radio,
  title={The Radio Fundamental Catalog. I. Astrometry},
  author={Petrov, LY and Kovalev, Yuri Y},
  journal={The Astrophysical Journal Supplement Series},
  volume={276},
  number={2},
  pages={38},
  year={2025},
  publisher={The American Astronomical Society}
}

@article{shannon2025commensal,
  title={The commensal real-time ASKAP fast transient incoherent-sum survey},
  author={Shannon, Ryan M and Bannister, Keith W and Bera, Apurba and Bhandari, Shivani and Day, Cherie K and Deller, Adam T and Dial, Tyson and Dobie, Dougal and Ekers, Ron D and Fong, Wen-fai and others},
  journal={Publications of the Astronomical Society of Australia},
  volume={42},
  pages={e036},
  year={2025},
  publisher={Cambridge University Press}
}


\end{document}